\documentclass[aps,preprintnumbers,amsmath,amssymb,prd,superscriptaddress,longbibliography,twocolumn]{revtex4-1}
\usepackage{dcolumn}
\usepackage{color}
\usepackage[caption=false]{subfig}
\usepackage{feynmp}
\usepackage{feynmp-auto}
\usepackage{systeme,mathtools}
\usepackage{easyReview}

\def\comment#1{}

\newcommand{\nc}{\newcommand}
\nc{\beq}{\begin{eqnarray}}
\nc{\eeq}{\end{eqnarray}}
\nc{\scs}{\scriptstyle}
\nc{\setval}{\fmfset{wiggly_len}{3mm} \fmfset{arrow_len}{1.5mm}
	\fmfset{arrow_ang}{13} \fmfset{dash_len}{1.5mm}\fmfpen{0.125mm}
	\fmfset{dot_size}{2thick}}

\usepackage{bm,latexsym,mathrsfs,enumerate,color}
\usepackage[mathcal]{euscript}
\usepackage[breaklinks=true,unicode=true,urlcolor = blue,colorlinks = true,citecolor = blue,linkcolor = blue]{hyperref}
\usepackage{graphicx}
\usepackage{todonotes}
\usepackage{wrapfig}
\usepackage{easyReview}
\usepackage{mathbbol}
\usepackage{stmaryrd}

\renewcommand{\vec}[1]{\bm{#1}}

\def\slashchar#1{\setbox0=\hbox{$#1$}           
	\dimen0=\wd0                                 
	\setbox1=\hbox{/} \dimen1=\wd1               
	\ifdim\dimen0>\dimen1                        
	\rlap{\hbox to \dimen0{\hfil/\hfil}}      
	#1                                        
	\else                                        
	\rlap{\hbox to \dimen1{\hfil$#1$\hfil}}   
	/                                         
	\fi}                                         %

\DeclareMathAlphabet\mathbfcal{OMS}{cmsy}{b}{n}
\DeclareSymbolFontAlphabet{\amsmathbb}{AMSb}%

\def\sigmab{{\mbox{\boldmath $\sigma$}}}

\begin{document}
	
\title{ Deconfined criticality and bosonization duality in easy-plane Chern-Simons two-dimensional antiferromagnets}

\author{Vira Shyta}
\affiliation{Institute for Theoretical Solid State Physics, IFW Dresden, Helmholtzstr. 20, 01069 Dresden, Germany}
\affiliation{KAU — Department of Theoretical and Mathematical Physics, Kyiv Academic University, 36 Vernadsky blvd., Kyiv 03142, Ukraine}

\author{Jeroen van den Brink}
\affiliation{Institute for Theoretical Solid State Physics, IFW Dresden, Helmholtzstr. 20, 01069 Dresden, Germany}
\affiliation{Institute for Theoretical Physics, TU Dresden, 01069 Dresden, Germany}
\affiliation{W\"urzburg-Dresden Cluster of Excellence ct.qmat}

\author{Flavio S. Nogueira}
\affiliation{Institute for Theoretical Solid State Physics, IFW Dresden, Helmholtzstr. 20, 01069 Dresden, Germany}

\begin{abstract}
Two-dimensional quantum systems with competing orders can feature a deconfined quantum critical point, yielding a continuous phase transition that is incompatible with the Landau-Ginzburg-Wilson scenario, predicting instead a first-order phase transition. This is caused by the LGW order parameter breaking up into new elementary excitations at the critical point.
Canonical candidates for deconfined quantum criticality are quantum antiferromagnets with competing magnetic orders, captured by the easy-plane CP$^1$ model.
A delicate issue however is that numerics indicates the easy-plane CP$^1$  antiferromagnet to exhibit a first-order transition.
Here we show that an additional topological Chern-Simons term in the action changes this picture completely in several ways.  
We find that the topological easy-plane antiferromagnet undergoes a second-order transition with quantized critical exponents. 
Further, a particle-vortex duality naturally maps the partition function of the Chern-Simons easy-plane antiferromagnet into one of {\it massless} Dirac {\it fermions}. 
\end{abstract}

\maketitle

{\it Introduction} --- 
It is well known that some quantum critical systems exhibit a phase structure evading the traditional Landau-Ginzburg-Wilson (LGW) 
theory of phase transitions \cite{Senthil1490,Senthil_PhysRevB.70.144407,Sachdev-book}. Typical examples are 
two-dimensional quantum systems with competing orders, like for instance antiferromagnetic (AF) and valence-bond solid (VBS) orders 
originating from general quantum spin models with $SU(2)$ symmetry 
\cite{Read-Sachdev_PhysRevB.42.4568,Sachdev-book}. The LGW scenario predicts a first-order phase 
transition for such a system. However, the interplay between emergent instanton 
excitations (i.e., spacetime magnetic monopoles)  and staggered Berry phases 
\cite{Read-Sachdev_PhysRevB.42.4568} causes the actual phase transition to become a second-order one, leading
in this way to a quantum critical point separating the AF and VBS phases. For similar reasons discussed  
in studies of the deconfinement transition in high-energy physics, this type of critical point has been dubbed a 
“deconfined quantum critical point” \cite{Senthil1490}.  At such a critical point, order parameters on both sides of the transition fall apart into  “elementary particles” called spinons and we speak of spinon deconfinement.  

A well studied effective theory in this context is the quantum $O(3)$ nonlinear sigma model (${\rm NL}\sigma{\rm M}$), 
\begin{equation}
\mathcal{L}_{{\rm NL}\sigma{\rm M}}=\frac{1}{2g}(\partial_\mu\vec{n})^2+\dots,
\end{equation}
where $\vec{n}^2=1$, supplemented by instanton-suppressing terms, here symbolically 
represented by ellipses 
\cite{Senthil1490,Senthil_PhysRevB.70.144407,Motrunich-Vishwanath_PhysRevB.70.075104,KUKLOV20061602,Kragset_PhysRevLett.97.247201,Nogueira_PhysRevB.76.220403}. Physically, the model is an effective theory of antiferromagnets capturing the long-distance interactions, and the unit vector  $\vec{n}$ is the direction of the magnetization. When tuning the coupling constant $g$, the system undergoes a quantum phase transition from an AF ordered phase to a paramagnetic phase 
separated by a critical coupling $g_c$. 
By means of the Hopf map, $\vec{n}=z_a^*\sigmab_{ab}z_b$, where 
$\sigmab=(\sigma_x,\sigma_y,\sigma_z)$ is a Pauli matrix vector, 
the $O(3)$ ${\rm NL}\sigma{\rm M}$ is shown to be equivalent to the CP$^1$ model, 
\begin{equation}
\label{Eq:NCCP1-1}
\mathcal{L}_{\rm CP1}=\frac{1}{g}\sum_{a=1,2}|(\partial_\mu-ia_\mu)z_a|^2+\dots, 
\end{equation}
where the constraint $|z_1|^2+|z_2|^2=1$ holds and 
the gauge field is an auxiliary field given by $a_\mu=(i/2)\sum_a(z_a^*\partial_\mu z_a-z_a\partial_\mu z_a^*)$. 

Although the gauge field $a_\mu$ is an auxiliary field at the level of field equations, it becomes dynamical when quantum 
fluctuations of the spinon fields $z_a$ are accounted for, causing a Maxwell term to be generated in the low-energy regime \cite{Hikami}. 
In this context it is also interesting to consider generalizations with $N$ complex fields, 
yielding an $O(2N)$ symmetric version, the CP$^{N-1}$ model. It has been recently 
demonstrated \cite{Pelissetto_PhysRevE.101.062136} that the large $N$ limit in a instanton-suppressed CP$^{N-1}$ model implies a second-order phase transition. The result agrees with the standard field theory analysis of the large $N$ limit \cite{Hikami,Nogueira_PhysRevB.86.045121}.  Nevertheless, lower values of $N$ were shown numerically to exhibit a first-order phase transition, specifically for $N=4, 10, 15$; though the $N=2$ case remained inconclusive  
\cite{Pelissetto_PhysRevE.101.062136,Smiseth_PhysRevB.71.214509}. This result contrasts with the large $N$ limit without instanton suppression, where a first-order phase transition 
occurs \cite{Nahum_PhysRevB.88.134411,Pelissetto_2020}. 

A well-studied model since the early days of DC \cite{Senthil1490,Senthil_PhysRevB.70.144407,KUKLOV20061602,Kragset_PhysRevLett.97.247201} 
is the easy-plane CP$^1$ model with Lagrangian,
\begin{equation}
		\label{Eq:NCCP1}
		\mathcal{L}_{\rm ep}=\mathcal{L}_{\rm M}+
		\mathcal{L}_{\rm CP1}+\frac{K}{2g^2}(|z_1|^2-|z_2|^2)^2,
\end{equation}
which follows directly from the ${\rm NL}\sigma{\rm M}$  by adding the easy-plane anisotropy term, 
$\mathcal{L}_{\rm anis}=Kn_z^2/2g^2$, where $K>0$. Instanton suppression 
in the above Lagrangian is achieved by means of a Maxwell term \cite{Motrunich-Vishwanath_PhysRevB.70.075104,Senthil_PhysRevB.70.144407}, 
 \begin{equation}
 	\label{eq:bareMaxwell}
 	\mathcal{L}_{\rm M}=\frac{1}{2 e^2}(\epsilon_{\mu\nu\lambda}\partial_\nu a_\lambda)^2.
 \end{equation}
 
An exact particle-vortex duality transformation of 
the lattice Villain model version of $\mathcal{L}_{\rm ep}$ 
shows that the model is self-dual 
\cite{Motrunich-Vishwanath_PhysRevB.70.075104,Senthil1490,Senthil_PhysRevB.70.144407,Smiseth_PhysRevB.71.214509}. 
Partly on the basis of this self-duality, it was originally argued \cite{Senthil1490,Senthil_PhysRevB.70.144407} that the easy-plane CP$^1$  
model undergoes a second-order phase transition, featuring therefore a deconfined quantum critical point. However, it was later 
demonstrated numerically 
that the phase transition is actually a first-order one \cite{KUKLOV20061602,Kragset_PhysRevLett.97.247201}, 
a result that is also corroborated by renormalization group (RG) results \cite{Nogueira_PhysRevB.76.220403}.

Here we consider the topological easy-plane CP$^1$ lagrangian including a Chern-Simons (CS) term, i.e.,  
$\mathcal{L}=\mathcal{L}_{\rm ep}+\mathcal{L}_{\rm CS}$, where, 
\begin{equation}
\label{Eq:L-CS}
\mathcal{L}_{\rm CS}=i\frac{\kappa}{2}\epsilon_{\mu\nu\lambda}a_\mu\partial_\nu a_\lambda, 
\end{equation} 
describes a CS Lagrangian in Euclidean spacetime. For arbitrary real $\kappa$ the CS action is invariant under any topologically trivial gauge transformation, since the surface term vanishes in this case. On the other hand, topologically nontrivial ones generate a surface term which does not vanish. In this case one demands the invariance of $\exp(-S_{\rm CS})$, which forces $\kappa$ to be quantized, $\kappa= n/(2\pi)$, where $n\in\mathbb{Z}$ is the CS level \cite{Witten_RevModPhys.88.035001,witten2016three}. 

The motivation for  such a system is twofold. Firstly, it is interesting to examine the case of the instanton suppression by a topological term instead of a  bare Maxwell term. Secondly, a system with similar properties should arise in the context of chiral spin liquids \cite{Wen-Wilczek-Zee_PhysRevB.39.11413}. 
Moreover, as we will elaborate later, this is of direct relevance to bilayer quantum Hall systems that have been realized experimentally.  

This paper consists of three parts. First, we perform an RG analysis of 
the CP$^1$ CS action and show that the fixed point structure implies a 
second-order phase transition with critical exponents depending on the CS coupling and, hence, forming a new universality class. We will see that the scaling behavior of the topological theory cannot be 
smoothly connected to the limit where $\kappa\to 0$. 
In the second part of the paper we 
show that the dual model features a CS term of  the form,  
\begin{equation}
\label{Eq:K-CS}
\widetilde{\mathcal{L}}_{CS}=-\frac{i}{2\kappa}\epsilon_{\mu\nu\lambda}(b_{1\mu}+b_{2\mu})\partial_\nu(b_{1\lambda}+b_{2\lambda}),
\end{equation} 
with two gauge fields $b_{1\mu}$ and $b_{2\mu}$. 
Finally, in the third part we show that  for $\kappa = 1/(2\pi)$ 
the duality of the second part actually corresponds to a 
bosonization duality \cite{SEIBERG2016395,Karch_PhysRevX.6.031043} 
involving {\it massless} Dirac fermions \cite{Senthil_PhysRevX.7.031051}.

{\it Renormalization group analysis} --- 
Let us start by discussing the nature of the phase transition of the easy-plane CP$^1$ CS model by means of RG calculations. In order to regularize the short distance behavior, 
we also include the Maxwell term \eqref{eq:bareMaxwell} in the Lagrangian $\mathcal{L}=\mathcal{L}_{\rm ep}+\mathcal{L}_{\rm CS}$, and consider a soft 
constraint version of the model, 
\begin{eqnarray}	
	\label{eq:CP1soft}
	\mathcal{L}&=&\mathcal{L}_M+\mathcal{L}_{\rm CS}+
	\sum_{a=1,2}\left[|(\partial_\mu-ia_\mu)z_a|^2+m_0^2|z_a|^2\right]\nonumber\\
	&+&\frac{u}{2}(|z_1|^2+|z_2|^2)^2
	+	\frac{K}{2}(|z_1|^2- |z_2|^2)^2.
\end{eqnarray}
Details of the RG calculations are presented in  Supplemental Material (SM). 
There we show that the original theory features two IR fixed points for the renormalized  dimensionless couplings $\hat{u}$, $\hat{K}$, and $\hat{e}^2$. Importantly, $e^2$ sets a UV scale for the renormalized dimensionless gauge coupling $\hat{e}^2$, in the sense that the IR stable fixed point $\hat{e}^2_*$ is also reached when $e^2\to\infty$ (see SM).    
	One of the fixed points is $O(2)\times O(2)$-symmetric, while the second one corresponds to an emergent $O(4)$-symmetry. Interestingly, the Abelian Higgs CS critical exponents do not belong to the $XY$ universality class, as they are $\kappa$-dependent.

An important outcome of the RG analysis is that the limit $\kappa\to 0$ with $e^2$ finite does not reduce to the RG equations expected 
for a $U(1)\times U(1)$ Abelian Higgs model \cite{Hikami}. This happens because  
the presence of the CS term causes the one-loop gauge field bubble in the scalar 
field vertex function to vanish at zero external momenta (see SM for details on this 
point). 
 
From the RG analysis it follows that the correlation length critical exponents for the $O(2)\times O(2)$-  and $O(4)$-symmetric IR fixed points are quantized and depend on the level of the CS term. In particular, for a level 1 CS term this yields $\nu^{O(2)\times O(2)}=49/80\approx 0.613$. This value 
is nearly the same as 
the one-loop result $\nu=5/8$ of the $XY$ universality class. For the $O(4)$-symmetric criticality we obtain a larger value, $\nu^{O(4)}=2/3$, which is independent of the CS level at the one-loop order. 

The anomalous dimension $\eta_N$ is defined by the critical 
magnetization correlation function at large distances, $\langle\vec{n}(x)\cdot\vec{n}(0)\rangle\sim 1/|x|^{1+\eta_N(n)}$. 
For a level 1 CS term we obtain, $	\eta_N^{O(2)\times O(2)}=59/49\approx 1.2$ and 
$\eta_N^{O(4)}=164/147\approx 1.12$, for the $O(2)\times O(2)$ and $O(4)$ symmetric cases, respectively. This clearly shows that a new universality class 
emerges.   

At this point the following remark is in order. Typically, DC implies 
considerably larger  
anomalous dimensions $\eta_N$ as compared to the case of the LGW paradigm of phase 
transitions. However, it is rather rare that these values exceed unity.  
The leading order value in the easy-plane case without a CS term is $\eta_N=1$ 
(Gaussian approximation) \cite{Senthil1490}. For the $J-Q$ model the result is 
$\eta_N\approx 0.35$, but the easy-plane $J-Q$ model is reported to deliver a 
much larger value, $\eta_N\approx 0.91$ \cite{Sandvik_PhysRevX.7.031052}. 
On the other hand, the theory considered here exhibits anomalous dimensions 
$\eta_N>1$. An example where this also occurs is in a lattice 
boson model with an emergent $Z_2$ gauge symmetry \cite{Isakov193}, where the anomalous dimension is numerically 
calculated to be $\eta\approx 1.493$.   


{\it Duality analysis} --- 
We start the discussion of the duality transformation by
 changing to polar coordinates $z_{a}=\rho_a e^{i\theta_a}$ in the partition function of the easy-plane CS CP$^1$ model. After integrating  $\rho_2$ out and assuming a strong anisotropy ($K\gg g^2$), we obtain 
$\rho_1^2\approx\rho_2^2\approx  1/2$, leading to an effective action depending only on the phase fields coupled to the gauge field, 
\begin{equation}
	\label{Eq:Seff}
	S_{\rm eff}=S_{\rm CS}+ \frac{1}{2g}\sum_{a=1,2}\int d^3x(\partial_\mu\theta_a-a_\mu)^2,
\end{equation}  
where the CS action $S_{\rm CS}$ corresponds to the Lagrangian (\ref{Eq:L-CS}). The above effective action is equivalent to 
a two-component CS superconductor in the London limit where the amplitudes of the order parameter are constrained to be equal.

The traditional way to perform a duality transformation 
	is to carry it out on the lattice \cite{kleinert1989gauge}.   
	Nevertheless, while it is a straightforward task to define a Maxwell term on the lattice \cite{Peskin1978},
fundamental difficulties arise when one tries to define the CS term on the lattice.
It is known to be problematic to enforce the properties of a topological continuum field theory consistently on the lattice 
\cite{Frohlich-Marchetti,ELIEZER1992118,BERRUTO2000366}, 
although recently considerable progress has been made \cite{Gu-Wen_PhysRevB.90.115141,Gaiotto-Kapustin,demarco2019lattice}.
For these reasons, we will restrict ourselves to performing the subsequent calculations directly in the continuum. 

Even though we are working directly in the continuum, in order 
	for the theory to be well-defined  at the short distances, we need to regularize 
	it. So  we include 
	an additional Maxwell term \cite{HanssonKarlhedeRocek}. 
	The first step of our duality transformation introduces auxiliary fields $h_{I\mu}$, $I=1,2$, such that, 
\begin{eqnarray}
	\label{Eq:MCS-Higgs}
S_{\rm eff}'&=&\sum_{I=1,2}\int d^3x\left[\frac{g}{2}h_{I\mu}^2-ih_{I\mu}(\partial_\mu\theta_I-a_\mu)\right]\\&+&\frac{1}{2e^2}\int d^3x(\epsilon_{\mu\nu\lambda}\partial_\nu a_\lambda)^2+
i\frac{\kappa}{2}\int d^3x \epsilon_{\mu\nu\lambda}a_\mu\partial_\nu a_\lambda.\nonumber
\end{eqnarray}
To account for the periodicity of $\theta_I$,
the following decomposition in terms of longitudinal phase fluctuations and vortex 
gauge fields holds \cite{kleinert1989gauge}, 
%
$\partial_\mu\theta_I=\partial_\mu\varphi_I+2\pi v_{I\mu}$,
where $\varphi_I\in\mathbb{R}$ and the vorticity,  
\begin{equation}
	\label{Eq:Vortex-field}
	w_{I\mu}=\epsilon_{\mu\nu\lambda}\partial_\nu v_{I\lambda}(x)=\sum_c n_{Ic}\oint_{L_{Ic}}dy_\mu^{(c)}\delta^3(x-y^{(c)}),
\end{equation}
with quanta $n_{Ic}\in\mathbb{Z}$ and the integral is over a path along the 
$c$-th vortex loop $L_{Ic}$.   

Integrating out 
both $\varphi_I$ and $a_\mu$ leads to the action, 
\begin{eqnarray}
	\label{eq:seffdmunu}
	\widetilde{S}&=&\sum_{I=1,2}\int d^3x\left(\frac{g}{2}h_{I\mu}^2+i2\pi v_{I\mu} h_{I\mu}\right)\\
	&+&\frac{1}{2}\int d^3x\int d^3x'D_{\mu\nu}(x-x')(h_{1\mu}+ h_{2\mu})(h'_{1\nu}+ h'_{2\nu}),\nonumber
\end{eqnarray}
where $h'_{I\mu}$ denotes dependence on $x'$, and the propagator in momentum space,
\begin{equation}
	D_{\mu\nu}(p)=\frac{e^2}{p^2+e^4\kappa^2}\left(\delta_{\mu\nu}-e^2\kappa
	\epsilon_{\mu\nu\lambda}\frac{p_\lambda}{p^2}\right),
\end{equation}
is the Fourier transform of $D_{\mu\nu}(x)$. Here, the longitudinal 
contribution is absent due to the constraint $\partial_\mu h_{I\mu}=0$ which appears after integrating out fields $\varphi_I$. This also leads to  $h_{I\mu}$ being expressed in terms of new auxiliary fields $b_{I\mu}$ as  $h_{I\mu}=\epsilon_{\mu\nu\lambda}\partial_\nu b_{I\lambda}$.

As we are interested in the case of easy-plane CS CP$^1$ model, we can send $e^2\to\infty$ after performing explicitly the calculations in Eq. \eqref{eq:seffdmunu} and obtain the following dual Lagrangian,
\begin{eqnarray}
	\label{Eq:Seff-dual}
	\mathcal{L}_{\text{dual}}&=&\sum_{I=1,2}\left[\frac{g}{2}(\epsilon_{\mu\nu\lambda}\partial_\nu b_{I \lambda})^2 	+i2\pi w_{I\mu} b_{I\mu}\right]\nonumber\\&-&\frac{i}{2\kappa}\epsilon_{\mu\nu\lambda}
	(b_{1\mu}+ b_{2\mu})\partial_\nu (b_{1\lambda}+b_{2\lambda}).
\end{eqnarray}
One notices that the presence of the CS term in the original model leads to the appearance of the mixed CS term anticipated in Eq. \eqref{Eq:K-CS}. 
Thus, the dual action \eqref{Eq:Seff-dual} features gauge fields coupled to an ensemble of vortex loops $w_{I\mu}$. The latter represent the 
worldlines of the particles of the original model \cite{Peskin1978,THOMAS1978513}.  

As mentioned earlier in the context of the original theory using a soft constraint, an IR stable fixed point for the dimensionless renormalized gauge coupling is reached as $e^2\to\infty$. This result remains valid in the hard constraint case. In Eq.  (\ref{Eq:Seff-dual}) $1/g$ assumes the role of $e^2$ of the original theory. Note that $g=\hat{g}/\Lambda$, where $\hat{g}$ is dimensionless and $\Lambda$ is a UV cutoff, so the theory with a hard constraint reaches a UV nontrivial fixed point $\hat{g}_*$ as $\Lambda\to\infty$, so $g\to 0$. Thus, the duality establishes a mapping between the UV and IR regimes of the theory.

{\it Bosonization duality} --- Having obtained a bosonic dual theory, we will show now that the theory of CS easy-plane antiferromagnets is actually self-dual at criticality and leads to the bosonization duality for massless Dirac fermions. We proceed to show this by first integrating out the fields $b_{I\mu}$ in Eq. \eqref{Eq:Seff-dual}. This yields the dual action in terms of vortex loop fields, 
 \begin{eqnarray}
 	\label{Eq:S-dual-vortex}
 	\widetilde{S}&=&2\pi^2\int d^3x\int d^3x'\widetilde{D}_{\mu\nu}(x-x')(w_{1\mu}+w_{2\mu})(w_{1\nu}'+w_{2\nu}')
 	\nonumber\\
 	&+&\frac{\pi}{g}\int d^3x\int d^3x'\frac{(w_{1\mu}-w_{2\mu})(w_{1\mu}'-w_{2\mu}')}{|x-x'|},
 \end{eqnarray}%
where as before we are using primes to denote the dependence on $x'$ and 
$\widetilde{D}_{\mu\nu}(x-x')$ in momentum space reads, 
%
%
%
\begin{equation}
	\widetilde{D}_{\mu \nu}(p)=\frac{g \kappa^2}{2\left(g^{2} \kappa^{2} p^{2}+4\right)}\left(\delta_{\mu \nu}-2\frac{\varepsilon_{\mu \nu \lambda}p_{\lambda}}{\kappa g p^{2}}\right).
\end{equation}
Now, we will show that, similarly to the standard easy-plane theory 
\cite{Motrunich-Vishwanath_PhysRevB.70.075104}, the model considered here is 
self-dual in the large distance regime $g^2p^2\ll 1$. In this case the vortices 
$w_{1\mu}$ and $w_{2\mu}$ balance, so we can write approximately, 
$w_{1\mu}=w_{2\mu}\equiv w_\mu$, so that (for details, see SM), 
\begin{equation}
	\label{eq:ReCdual-lag-one-field}
	S_{\text{dual}}=\int d^3x\left(2\pi^2 g \kappa^2w_{\mu}^2+i2\pi^2\kappa v_\mu w_\mu\right).
\end{equation}
On the other hand, letting  
$g\to 0$ in the initial Abelian Higgs CS action  (\ref{Eq:MCS-Higgs}) 
and integrating out $h_{2\mu}$ yields $a_\mu=\partial_\mu\theta_2$. Subsequent 
integration of $h_{1\mu}$ enforces $\theta_1=\theta_2\equiv\theta$. 
At the end, this yields,
\begin{equation}
	\label{eq:ReCinitial-lag-one-field}
	S=\int d^3x\left(\frac{2\pi^2}{e^2}w_\mu^2+i2\pi^2\kappa v_\mu w_\mu
	\right),
\end{equation}
and therefore we obtain the duality for the partition function, 
\begin{equation}
	\label{Eq:self-dual}
		Z_{\text{dual}}(e^2=\infty,g,\kappa)=
		Z(g'=0,e'^2=1/(g\kappa^2),\kappa).
\end{equation}
Underlying the above result is the duality relation between the couplings, 
$ge^2=1/\kappa^2$. For a level 1 CS term the latter reduces to $g e^2 = (2\pi)^2$, which is the Dirac quantization associated to particle-vortex duality. It is interesting to note that Eq. (\ref{Eq:self-dual}) constitutes a topological version of the "frozen superconductor" regime in the particle-vortex duality for the Abelian Higgs 
model in 2+1 dimensions derived by Peskin \cite{Peskin1978} and 
Dasgupta and Halperin \cite{Dasgupta-Halperin_PhysRevLett.47.1556}.


We are now ready to explore the critical dual theory which, as was discussed above, is obtained by setting $g\to 0$ in the Lagrangian \eqref{eq:ReCdual-lag-one-field}. This yields up to an overall normalization the partition function, 
%
\begin{eqnarray}
	\label{Eq:Z-tilde-0}
	\widetilde{Z}_{\rm crit}&=&\sum_{\rm loops}\exp\left[i\frac{\pi\kappa}{2}\sum_{a,b}n_an_b
	\right. \nonumber\\
	&\times&\left.\oint_{L_a}dx_\mu^{(a)}\oint_{L_b}dx_\nu^{(b)}
	\epsilon_{\mu\nu\lambda}\frac{(x^{(b)}-x^{(a)})_\lambda}{|x^{(b)}-x^{(a)}|^3}\right], 
\end{eqnarray}
where we sum over all loops $L_a$ and $L_b$, not excluding $a=b$ contributions, which will 
turn out to be a crucial point \cite{Polyakov-1988,turker2020bosonization}. 
For $a\neq b$ the double integral above yields a contribution $e^{i2\pi^2N_{ab}\kappa}$, $N_{ab}\in\mathbb{Z}$, in virtue of the Gauss linking number 
formula \cite{cf5mathematischen,frankel2011geometry}. 
Despite looking at first sight singular, the $a=b$ contributions are actually finite and proportional to the so called 
writhe of the (vortex) loop \cite{calugareanu1959integrale,calugareanu1961classes,White}. The latter can be conveniently written in terms of a 
suitable parametrization, $x_\mu(s)$, $s\in[0,1]$, by defining the unit vector, $u_\mu(s,s')=(x_\mu(s)-x_\mu(s'))/|x(s)-x(s')|$, in which case the 
writhe is recast as, 
\begin{eqnarray}
	\label{Eq:Writhe}
	\mathcal{W}_a&=&\frac{1}{4\pi}\int_{L_a}ds\int_{L_a}ds'\epsilon_{\mu\nu\lambda}\frac{dx_\mu}{ds}\frac{dx_\nu}{ds'}
	\frac{(x_\lambda(s)-x_\lambda(s'))}{|x(s)-x(s')|^3}
	\nonumber\\
	&=&\frac{1}{4\pi}\int_{L_a}ds\int_{L_a}ds'\epsilon_{\mu\nu\lambda}u_\mu\partial_su_\nu\partial_{s'}u_\lambda.
\end{eqnarray}
  The result is reminiscent of the point-splitting regularization employed to calculate expectation values of Wilson loops \cite{Witten_Jones}. This is in agreement with Ref. \cite{HanssonKarlhedeRocek}, where it is shown that the point-splitting procedure yields the topological invariant which coincides with the writhe in theories containing a Maxwell term in addition to a CS one when $e^2\to \infty$. 

We now consider a specific case of a level 1 CS theory in the original model corresponding to  	$\kappa=1/(2\pi)$. Consequently, the dual partition function at criticality  (\ref{Eq:Z-tilde-0}) takes the form,
\begin{equation}
	\label{Eq:Z-tilde-0-1}
	\widetilde{Z}_{\rm crit}=\sum_{\rm loops}(-1)^{N_{ab}}e^{i\pi \sum_{a}n_a^2\mathcal{W}_a}.
\end{equation}
The contribution from the linking number formula generates weight factors $(-1)^{n}$ in the dual model, where $n$ is integer. This result is reminiscent of 	the lack of gauge invariance of the partition function under topologically nontrivial gauge transformations in the dual model
	\cite{Redlich_PhysRevLett.52.18,Redlich_PhysRevD.29.2366}.  This result makes apparent 
	that the considered duality corresponds to a form of bosonization akin to the one discussed 
	by Polyakov for the CP$^1$ model with a CS term \cite{Polyakov-1988,FERREIROS20181}. This contribution is sometimes 
referred to as the Polyakov spin factor \cite{polyakov1990two,Polyakov-1988,AMBJORN1990509,Grundberg1990,Goldman2018,turker2020bosonization}.  Equation (\ref{Eq:Z-tilde-0-1}) relates to the representation of the partition function of a Dirac 
fermion in 2+1 Euclidean dimensions in terms of loops \cite{turker2020bosonization,Grundberg1990,AMBJORN1990509,Goldman2018}, with the difference 
that in our case the parity anomaly factor implies that the fermions are massless 
\cite{WITTEN1982324,Witten_RevModPhys.88.035001,ALVAREZGAUME1985288,FORTE1987252}.   

As far as the writhe is concerned, it is worth to recall that it 
arises quite naturally in the partition function of Wilson fermions on an euclidean cubic spacetime lattice \cite{turker2020bosonization}. 
However, the analysis of Ref. \cite{turker2020bosonization} and previous ones \cite{polyakov1990two,Polyakov-1988,AMBJORN1990509,Grundberg1990,Goldman2018, Raghu_PhysRevLett.120.016602} 
requires massive fermions.   

It is remarkable that even if the analysis above does not explicitly employ fermions, still a result that can only follow from massless fermions is obtained. To elaborate this point further we recall that a topologically nontrivial 
gauge transformation $\gamma$, $a_\mu\to a_\mu^\gamma$, in a continuous deformation of the gauge field, leads to the subsequent transformation of the fermion determinant 
$\det(\slashchar{\partial}+i\slashchar{a})\to (-1)^n \det(\slashchar{\partial}+i\slashchar{a}^\gamma)$, with $n$ being the winding number 
\cite{WITTEN1982324,ALVAREZGAUME1985288,Redlich_PhysRevD.29.2366,FORTE1987252}. Therefore, integrating over $a_\mu$ requires to 
account for redundant gauge 
configurations and sum over all possible winding numbers corresponding to 
different topological sectors in the partition function. 

To further substantiate our bosonization claim, we rederived this result using the flux attachment approach to duality \cite{Karch_PhysRevX.6.031043}, which involves a path integral formalism corresponding to a “Fourier transform” for quantized fluxes. In order for this to work in our case we have to attach fluxes to both fermions and bosons. The end result is that the dual Lagrangian (\ref{Eq:Seff-dual}) is the bosonized version of massless Dirac fermions with half-quantized CS flux attached. (The explicit derivation can be found in the SM). Therefore, our derivation is  consistent with the flux attachment technique, but in  contrast to it, does not assume any conjectures as a starting point. Thus, our analysis provides yet a further check for these conjectures.


{\it Final remarks} --- We have demonstrated through RG analysis that the topological easy-plane CP$^1$ model undergoes a second-order phase transition. Following this result, we established a dual theory, which at criticality exhibits a parity anomaly. This occurs at the particular value of a CS coupling $\kappa$ that provides topological gauge invariance. We relate that to massless Dirac fermions, thereby establishing an explicit bosonization duality \cite{SEIBERG2016395}. Since the theory we consider here possesses a $U(1)\times U(1)$ symmetry, 
our analysis subscribes into the so called beyond flavor bound scenario of duality 
\cite{Aharony-JHEP-2016,Hsin-Seiberg}. 

Additionally, let us consider these results within an experimental context. 
The dual theory (\ref{Eq:Seff-dual}) with  $\kappa=1/(2\pi)$ and gauge fields rescaled as $b_{I\mu}\rightarrow  b_{I\mu}/(2\pi)$ features a CS term 
as it occurs in  the $(1,1,1)$ quantum Hall (QH) state associated to a bilayer QH system \cite{Wen-Zee_PhysRevLett.69.1811,wen2004quantum,Kim_PhysRevB.63.205315}. As mentioned, the  initial model 
corresponds to a two-component CS superconductor. Therefore,  
the duality picture discussed here naturally connects 
the observed resonant tunneling in bilayer QH ferromagnets \cite{Spielman_PhysRevLett.84.5808} to a Josephson-like effect 
in a system that is not superconducting \cite{Wilczek_PhysRevLett.86.1833,Balents_PhysRevLett.86.1825,Stern_PhysRevLett.86.1829}. Our analysis shows that such an experimental setup represents the dual physical system to the actual easy-plane CS antiferromagnet. They belong to the same universality class so that the bilayer QH ferromagnet offers a controllable experimental system for a deconfined critical point. Moreover, in view of the connection to massless Dirac fermions established in this letter, bilayer QH ferromagnets would in principle offer a 
platform to experimentally explore the bosonization duality in 2+1 dimensions. It would be interesting to check whether experiments can reveal the critical behavior with quantized exponents as we predict here. 

Another system of interest where our approach may (with appropriate modifications) be relevant is the topological field theory for magic-angle graphene \cite{Vishwanath-tbg}, where a duality between superconductivity and insulating regimes occur.

\begin{acknowledgments}
	
	We thank the DFG for support through the W\"urzburg-Dresden Cluster of Excellence on Complexity and Topology in Quantum Matter – ct.qmat (EXC 2147, project-id 39085490) and through SFB 1143 (project-id 247310070). V.S. has been supported by UKRATOP-project (funded by BMBF with  
	grant number 01DK18002). 
\end{acknowledgments}

\bibliography{dqc+cs-refs-with-SM} 

\begin{thebibliography}{66}%
\makeatletter
\providecommand \@ifxundefined [1]{%
 \@ifx{#1\undefined}
}%
\providecommand \@ifnum [1]{%
 \ifnum #1\expandafter \@firstoftwo
 \else \expandafter \@secondoftwo
 \fi
}%
\providecommand \@ifx [1]{%
 \ifx #1\expandafter \@firstoftwo
 \else \expandafter \@secondoftwo
 \fi
}%
\providecommand \natexlab [1]{#1}%
\providecommand \enquote  [1]{``#1''}%
\providecommand \bibnamefont  [1]{#1}%
\providecommand \bibfnamefont [1]{#1}%
\providecommand \citenamefont [1]{#1}%
\providecommand \href@noop [0]{\@secondoftwo}%
\providecommand \href [0]{\begingroup \@sanitize@url \@href}%
\providecommand \@href[1]{\@@startlink{#1}\@@href}%
\providecommand \@@href[1]{\endgroup#1\@@endlink}%
\providecommand \@sanitize@url [0]{\catcode `\\12\catcode `\$12\catcode
  `\&12\catcode `\#12\catcode `\^12\catcode `\_12\catcode `\%12\relax}%
\providecommand \@@startlink[1]{}%
\providecommand \@@endlink[0]{}%
\providecommand \url  [0]{\begingroup\@sanitize@url \@url }%
\providecommand \@url [1]{\endgroup\@href {#1}{\urlprefix }}%
\providecommand \urlprefix  [0]{URL }%
\providecommand \Eprint [0]{\href }%
\providecommand \doibase [0]{http://dx.doi.org/}%
\providecommand \selectlanguage [0]{\@gobble}%
\providecommand \bibinfo  [0]{\@secondoftwo}%
\providecommand \bibfield  [0]{\@secondoftwo}%
\providecommand \translation [1]{[#1]}%
\providecommand \BibitemOpen [0]{}%
\providecommand \bibitemStop [0]{}%
\providecommand \bibitemNoStop [0]{.\EOS\space}%
\providecommand \EOS [0]{\spacefactor3000\relax}%
\providecommand \BibitemShut  [1]{\csname bibitem#1\endcsname}%
\let\auto@bib@innerbib\@empty
\bibitem [{\citenamefont {Senthil}\ \emph
  {et~al.}(2004{\natexlab{a}})\citenamefont {Senthil}, \citenamefont
  {Vishwanath}, \citenamefont {Balents}, \citenamefont {Sachdev},\ and\
  \citenamefont {Fisher}}]{Senthil1490}%
  \BibitemOpen
  \bibfield  {author} {\bibinfo {author} {\bibfnamefont {T.}~\bibnamefont
  {Senthil}}, \bibinfo {author} {\bibfnamefont {Ashvin}\ \bibnamefont
  {Vishwanath}}, \bibinfo {author} {\bibfnamefont {Leon}\ \bibnamefont
  {Balents}}, \bibinfo {author} {\bibfnamefont {Subir}\ \bibnamefont
  {Sachdev}}, \ and\ \bibinfo {author} {\bibfnamefont {Matthew P.~A.}\
  \bibnamefont {Fisher}},\ }\bibfield  {title} {\enquote {\bibinfo {title}
  {Deconfined quantum critical points},}\ }\href {\doibase
  10.1126/science.1091806} {\bibfield  {journal} {\bibinfo  {journal}
  {Science}\ }\textbf {\bibinfo {volume} {303}},\ \bibinfo {pages} {1490--1494}
  (\bibinfo {year} {2004}{\natexlab{a}})}\BibitemShut {NoStop}%
\bibitem [{\citenamefont {Senthil}\ \emph
  {et~al.}(2004{\natexlab{b}})\citenamefont {Senthil}, \citenamefont {Balents},
  \citenamefont {Sachdev}, \citenamefont {Vishwanath},\ and\ \citenamefont
  {Fisher}}]{Senthil_PhysRevB.70.144407}%
  \BibitemOpen
  \bibfield  {author} {\bibinfo {author} {\bibfnamefont {T.}~\bibnamefont
  {Senthil}}, \bibinfo {author} {\bibfnamefont {Leon}\ \bibnamefont {Balents}},
  \bibinfo {author} {\bibfnamefont {Subir}\ \bibnamefont {Sachdev}}, \bibinfo
  {author} {\bibfnamefont {Ashvin}\ \bibnamefont {Vishwanath}}, \ and\ \bibinfo
  {author} {\bibfnamefont {Matthew P.~A.}\ \bibnamefont {Fisher}},\ }\bibfield
  {title} {\enquote {\bibinfo {title} {Quantum criticality beyond the
  landau-ginzburg-wilson paradigm},}\ }\href {\doibase
  10.1103/PhysRevB.70.144407} {\bibfield  {journal} {\bibinfo  {journal} {Phys.
  Rev. B}\ }\textbf {\bibinfo {volume} {70}},\ \bibinfo {pages} {144407}
  (\bibinfo {year} {2004}{\natexlab{b}})}\BibitemShut {NoStop}%
\bibitem [{\citenamefont {Sachdev}(2011)}]{Sachdev-book}%
  \BibitemOpen
  \bibfield  {author} {\bibinfo {author} {\bibfnamefont {S.}~\bibnamefont
  {Sachdev}},\ }\href@noop {} {\emph {\bibinfo {title} {Quantum Phase
  Transitions}}},\ \bibinfo {edition} {2nd}\ ed.\ (\bibinfo  {publisher}
  {Cambridge University Press},\ \bibinfo {year} {2011})\BibitemShut {NoStop}%
\bibitem [{\citenamefont {Read}\ and\ \citenamefont
  {Sachdev}(1990)}]{Read-Sachdev_PhysRevB.42.4568}%
  \BibitemOpen
  \bibfield  {author} {\bibinfo {author} {\bibfnamefont {N.}~\bibnamefont
  {Read}}\ and\ \bibinfo {author} {\bibfnamefont {Subir}\ \bibnamefont
  {Sachdev}},\ }\bibfield  {title} {\enquote {\bibinfo {title} {Spin-peierls,
  valence-bond solid, and n\'eel ground states of low-dimensional quantum
  antiferromagnets},}\ }\href {\doibase 10.1103/PhysRevB.42.4568} {\bibfield
  {journal} {\bibinfo  {journal} {Phys. Rev. B}\ }\textbf {\bibinfo {volume}
  {42}},\ \bibinfo {pages} {4568--4589} (\bibinfo {year} {1990})}\BibitemShut
  {NoStop}%
\bibitem [{\citenamefont {Motrunich}\ and\ \citenamefont
  {Vishwanath}(2004)}]{Motrunich-Vishwanath_PhysRevB.70.075104}%
  \BibitemOpen
  \bibfield  {author} {\bibinfo {author} {\bibfnamefont {Olexei~I.}\
  \bibnamefont {Motrunich}}\ and\ \bibinfo {author} {\bibfnamefont {Ashvin}\
  \bibnamefont {Vishwanath}},\ }\bibfield  {title} {\enquote {\bibinfo {title}
  {Emergent photons and transitions in the $\mathrm{O}(3)$ sigma model with
  hedgehog suppression},}\ }\href {\doibase 10.1103/PhysRevB.70.075104}
  {\bibfield  {journal} {\bibinfo  {journal} {Phys. Rev. B}\ }\textbf {\bibinfo
  {volume} {70}},\ \bibinfo {pages} {075104} (\bibinfo {year}
  {2004})}\BibitemShut {NoStop}%
\bibitem [{\citenamefont {Kuklov}\ \emph {et~al.}(2006)\citenamefont {Kuklov},
  \citenamefont {Prokof'ev}, \citenamefont {Svistunov},\ and\ \citenamefont
  {Troyer}}]{KUKLOV20061602}%
  \BibitemOpen
  \bibfield  {author} {\bibinfo {author} {\bibfnamefont {A.B.}\ \bibnamefont
  {Kuklov}}, \bibinfo {author} {\bibfnamefont {N.V.}\ \bibnamefont
  {Prokof'ev}}, \bibinfo {author} {\bibfnamefont {B.V.}\ \bibnamefont
  {Svistunov}}, \ and\ \bibinfo {author} {\bibfnamefont {M.}~\bibnamefont
  {Troyer}},\ }\bibfield  {title} {\enquote {\bibinfo {title} {Deconfined
  criticality, runaway flow in the two-component scalar electrodynamics and
  weak first-order superfluid-solid transitions},}\ }\href {\doibase
  https://doi.org/10.1016/j.aop.2006.04.007} {\bibfield  {journal} {\bibinfo
  {journal} {Annals of Physics}\ }\textbf {\bibinfo {volume} {321}},\ \bibinfo
  {pages} {1602 -- 1621} (\bibinfo {year} {2006})},\ \bibinfo {note} {july 2006
  Special Issue}\BibitemShut {NoStop}%
\bibitem [{\citenamefont {Kragset}\ \emph {et~al.}(2006)\citenamefont
  {Kragset}, \citenamefont {Sm\o{}rgrav}, \citenamefont {Hove}, \citenamefont
  {Nogueira},\ and\ \citenamefont {Sudb\o{}}}]{Kragset_PhysRevLett.97.247201}%
  \BibitemOpen
  \bibfield  {author} {\bibinfo {author} {\bibfnamefont {S.}~\bibnamefont
  {Kragset}}, \bibinfo {author} {\bibfnamefont {E.}~\bibnamefont
  {Sm\o{}rgrav}}, \bibinfo {author} {\bibfnamefont {J.}~\bibnamefont {Hove}},
  \bibinfo {author} {\bibfnamefont {F.~S.}\ \bibnamefont {Nogueira}}, \ and\
  \bibinfo {author} {\bibfnamefont {A.}~\bibnamefont {Sudb\o{}}},\ }\bibfield
  {title} {\enquote {\bibinfo {title} {First-order phase transition in
  easy-plane quantum antiferromagnets},}\ }\href {\doibase
  10.1103/PhysRevLett.97.247201} {\bibfield  {journal} {\bibinfo  {journal}
  {Phys. Rev. Lett.}\ }\textbf {\bibinfo {volume} {97}},\ \bibinfo {pages}
  {247201} (\bibinfo {year} {2006})}\BibitemShut {NoStop}%
\bibitem [{\citenamefont {Nogueira}\ \emph {et~al.}(2007)\citenamefont
  {Nogueira}, \citenamefont {Kragset},\ and\ \citenamefont
  {Sudb\o{}}}]{Nogueira_PhysRevB.76.220403}%
  \BibitemOpen
  \bibfield  {author} {\bibinfo {author} {\bibfnamefont {F.~S.}\ \bibnamefont
  {Nogueira}}, \bibinfo {author} {\bibfnamefont {S.}~\bibnamefont {Kragset}}, \
  and\ \bibinfo {author} {\bibfnamefont {A.}~\bibnamefont {Sudb\o{}}},\
  }\bibfield  {title} {\enquote {\bibinfo {title} {Quantum critical scaling
  behavior of deconfined spinons},}\ }\href {\doibase
  10.1103/PhysRevB.76.220403} {\bibfield  {journal} {\bibinfo  {journal} {Phys.
  Rev. B}\ }\textbf {\bibinfo {volume} {76}},\ \bibinfo {pages} {220403(R)}
  (\bibinfo {year} {2007})}\BibitemShut {NoStop}%
\bibitem [{\citenamefont {Hikami}(1979)}]{Hikami}%
  \BibitemOpen
  \bibfield  {author} {\bibinfo {author} {\bibfnamefont {Shinobu}\ \bibnamefont
  {Hikami}},\ }\bibfield  {title} {\enquote {\bibinfo {title} {{Renormalization
  Group Functions of CP(N-1) Non-Linear sigma-Model and N-Component Scalar QED
  Model}},}\ }\href {\doibase 10.1143/PTP.62.226} {\bibfield  {journal}
  {\bibinfo  {journal} {Progress of Theoretical Physics}\ }\textbf {\bibinfo
  {volume} {62}},\ \bibinfo {pages} {226--233} (\bibinfo {year}
  {1979})}\BibitemShut {NoStop}%
\bibitem [{\citenamefont {Pelissetto}\ and\ \citenamefont
  {Vicari}(2020{\natexlab{a}})}]{Pelissetto_PhysRevE.101.062136}%
  \BibitemOpen
  \bibfield  {author} {\bibinfo {author} {\bibfnamefont {Andrea}\ \bibnamefont
  {Pelissetto}}\ and\ \bibinfo {author} {\bibfnamefont {Ettore}\ \bibnamefont
  {Vicari}},\ }\bibfield  {title} {\enquote {\bibinfo {title}
  {Three-dimensional monopole-free ${\mathrm{cp}}^{N\ensuremath{-}1}$
  models},}\ }\href {\doibase 10.1103/PhysRevE.101.062136} {\bibfield
  {journal} {\bibinfo  {journal} {Phys. Rev. E}\ }\textbf {\bibinfo {volume}
  {101}},\ \bibinfo {pages} {062136} (\bibinfo {year}
  {2020}{\natexlab{a}})}\BibitemShut {NoStop}%
\bibitem [{\citenamefont {Nogueira}\ and\ \citenamefont
  {Sudb\o{}}(2012)}]{Nogueira_PhysRevB.86.045121}%
  \BibitemOpen
  \bibfield  {author} {\bibinfo {author} {\bibfnamefont {Flavio~S.}\
  \bibnamefont {Nogueira}}\ and\ \bibinfo {author} {\bibfnamefont {Asle}\
  \bibnamefont {Sudb\o{}}},\ }\bibfield  {title} {\enquote {\bibinfo {title}
  {Deconfined quantum criticality and logarithmic violations of scaling from
  emergent gauge symmetry},}\ }\href {\doibase 10.1103/PhysRevB.86.045121}
  {\bibfield  {journal} {\bibinfo  {journal} {Phys. Rev. B}\ }\textbf {\bibinfo
  {volume} {86}},\ \bibinfo {pages} {045121} (\bibinfo {year}
  {2012})}\BibitemShut {NoStop}%
\bibitem [{\citenamefont {Smiseth}\ \emph {et~al.}(2005)\citenamefont
  {Smiseth}, \citenamefont {Sm\o{}rgrav}, \citenamefont {Babaev},\ and\
  \citenamefont {Sudb\o{}}}]{Smiseth_PhysRevB.71.214509}%
  \BibitemOpen
  \bibfield  {author} {\bibinfo {author} {\bibfnamefont {J.}~\bibnamefont
  {Smiseth}}, \bibinfo {author} {\bibfnamefont {E.}~\bibnamefont
  {Sm\o{}rgrav}}, \bibinfo {author} {\bibfnamefont {E.}~\bibnamefont {Babaev}},
  \ and\ \bibinfo {author} {\bibfnamefont {A.}~\bibnamefont {Sudb\o{}}},\
  }\bibfield  {title} {\enquote {\bibinfo {title} {Field- and
  temperature-induced topological phase transitions in the three-dimensional
  $n$-component london superconductor},}\ }\href {\doibase
  10.1103/PhysRevB.71.214509} {\bibfield  {journal} {\bibinfo  {journal} {Phys.
  Rev. B}\ }\textbf {\bibinfo {volume} {71}},\ \bibinfo {pages} {214509}
  (\bibinfo {year} {2005})}\BibitemShut {NoStop}%
\bibitem [{\citenamefont {Nahum}\ \emph {et~al.}(2013)\citenamefont {Nahum},
  \citenamefont {Chalker}, \citenamefont {Serna}, \citenamefont {Ortu\~no},\
  and\ \citenamefont {Somoza}}]{Nahum_PhysRevB.88.134411}%
  \BibitemOpen
  \bibfield  {author} {\bibinfo {author} {\bibfnamefont {Adam}\ \bibnamefont
  {Nahum}}, \bibinfo {author} {\bibfnamefont {J.~T.}\ \bibnamefont {Chalker}},
  \bibinfo {author} {\bibfnamefont {P.}~\bibnamefont {Serna}}, \bibinfo
  {author} {\bibfnamefont {M.}~\bibnamefont {Ortu\~no}}, \ and\ \bibinfo
  {author} {\bibfnamefont {A.~M.}\ \bibnamefont {Somoza}},\ }\bibfield  {title}
  {\enquote {\bibinfo {title} {Phase transitions in three-dimensional loop
  models and the $c{P}^{n\ensuremath{-}1}$ sigma model},}\ }\href {\doibase
  10.1103/PhysRevB.88.134411} {\bibfield  {journal} {\bibinfo  {journal} {Phys.
  Rev. B}\ }\textbf {\bibinfo {volume} {88}},\ \bibinfo {pages} {134411}
  (\bibinfo {year} {2013})}\BibitemShut {NoStop}%
\bibitem [{\citenamefont {Pelissetto}\ and\ \citenamefont
  {Vicari}(2020{\natexlab{b}})}]{Pelissetto_2020}%
  \BibitemOpen
  \bibfield  {author} {\bibinfo {author} {\bibfnamefont {Andrea}\ \bibnamefont
  {Pelissetto}}\ and\ \bibinfo {author} {\bibfnamefont {Ettore}\ \bibnamefont
  {Vicari}},\ }\bibfield  {title} {\enquote {\bibinfo {title} {Large-n behavior
  of three-dimensional lattice {CP} n-1 models},}\ }\href {\doibase
  10.1088/1742-5468/ab7747} {\bibfield  {journal} {\bibinfo  {journal} {Journal
  of Statistical Mechanics: Theory and Experiment}\ }\textbf {\bibinfo {volume}
  {2020}},\ \bibinfo {pages} {033209} (\bibinfo {year}
  {2020}{\natexlab{b}})}\BibitemShut {NoStop}%
\bibitem [{\citenamefont
  {Witten}(2016{\natexlab{a}})}]{Witten_RevModPhys.88.035001}%
  \BibitemOpen
  \bibfield  {author} {\bibinfo {author} {\bibfnamefont {Edward}\ \bibnamefont
  {Witten}},\ }\bibfield  {title} {\enquote {\bibinfo {title} {Fermion path
  integrals and topological phases},}\ }\href {\doibase
  10.1103/RevModPhys.88.035001} {\bibfield  {journal} {\bibinfo  {journal}
  {Rev. Mod. Phys.}\ }\textbf {\bibinfo {volume} {88}},\ \bibinfo {pages}
  {035001} (\bibinfo {year} {2016}{\natexlab{a}})}\BibitemShut {NoStop}%
\bibitem [{\citenamefont {Witten}(2016{\natexlab{b}})}]{witten2016three}%
  \BibitemOpen
  \bibfield  {author} {\bibinfo {author} {\bibfnamefont {Edward}\ \bibnamefont
  {Witten}},\ }\bibfield  {title} {\enquote {\bibinfo {title} {Three lectures
  on topological phases of matter},}\ }\href {\doibase
  10.1393/ncr/i2016-10125-3} {\bibfield  {journal} {\bibinfo  {journal} {La
  Rivista del Nuovo Cimento}\ }\textbf {\bibinfo {volume} {39}},\ \bibinfo
  {pages} {313--370} (\bibinfo {year} {2016}{\natexlab{b}})}\BibitemShut
  {NoStop}%
\bibitem [{\citenamefont {Wen}\ \emph {et~al.}(1989)\citenamefont {Wen},
  \citenamefont {Wilczek},\ and\ \citenamefont
  {Zee}}]{Wen-Wilczek-Zee_PhysRevB.39.11413}%
  \BibitemOpen
  \bibfield  {author} {\bibinfo {author} {\bibfnamefont {X.~G.}\ \bibnamefont
  {Wen}}, \bibinfo {author} {\bibfnamefont {Frank}\ \bibnamefont {Wilczek}}, \
  and\ \bibinfo {author} {\bibfnamefont {A.}~\bibnamefont {Zee}},\ }\bibfield
  {title} {\enquote {\bibinfo {title} {Chiral spin states and
  superconductivity},}\ }\href {\doibase 10.1103/PhysRevB.39.11413} {\bibfield
  {journal} {\bibinfo  {journal} {Phys. Rev. B}\ }\textbf {\bibinfo {volume}
  {39}},\ \bibinfo {pages} {11413--11423} (\bibinfo {year} {1989})}\BibitemShut
  {NoStop}%
\bibitem [{\citenamefont {Seiberg}\ \emph {et~al.}(2016)\citenamefont
  {Seiberg}, \citenamefont {Senthil}, \citenamefont {Wang},\ and\ \citenamefont
  {Witten}}]{SEIBERG2016395}%
  \BibitemOpen
  \bibfield  {author} {\bibinfo {author} {\bibfnamefont {Nathan}\ \bibnamefont
  {Seiberg}}, \bibinfo {author} {\bibfnamefont {T.}~\bibnamefont {Senthil}},
  \bibinfo {author} {\bibfnamefont {Chong}\ \bibnamefont {Wang}}, \ and\
  \bibinfo {author} {\bibfnamefont {Edward}\ \bibnamefont {Witten}},\
  }\bibfield  {title} {\enquote {\bibinfo {title} {A duality web in 2+1
  dimensions and condensed matter physics},}\ }\href {\doibase
  https://doi.org/10.1016/j.aop.2016.08.007} {\bibfield  {journal} {\bibinfo
  {journal} {Annals of Physics}\ }\textbf {\bibinfo {volume} {374}},\ \bibinfo
  {pages} {395 -- 433} (\bibinfo {year} {2016})}\BibitemShut {NoStop}%
\bibitem [{\citenamefont {Karch}\ and\ \citenamefont
  {Tong}(2016)}]{Karch_PhysRevX.6.031043}%
  \BibitemOpen
  \bibfield  {author} {\bibinfo {author} {\bibfnamefont {Andreas}\ \bibnamefont
  {Karch}}\ and\ \bibinfo {author} {\bibfnamefont {David}\ \bibnamefont
  {Tong}},\ }\bibfield  {title} {\enquote {\bibinfo {title} {Particle-vortex
  duality from 3d bosonization},}\ }\href {\doibase 10.1103/PhysRevX.6.031043}
  {\bibfield  {journal} {\bibinfo  {journal} {Phys. Rev. X}\ }\textbf {\bibinfo
  {volume} {6}},\ \bibinfo {pages} {031043} (\bibinfo {year}
  {2016})}\BibitemShut {NoStop}%
\bibitem [{\citenamefont {Wang}\ \emph {et~al.}(2017)\citenamefont {Wang},
  \citenamefont {Nahum}, \citenamefont {Metlitski}, \citenamefont {Xu},\ and\
  \citenamefont {Senthil}}]{Senthil_PhysRevX.7.031051}%
  \BibitemOpen
  \bibfield  {author} {\bibinfo {author} {\bibfnamefont {Chong}\ \bibnamefont
  {Wang}}, \bibinfo {author} {\bibfnamefont {Adam}\ \bibnamefont {Nahum}},
  \bibinfo {author} {\bibfnamefont {Max~A.}\ \bibnamefont {Metlitski}},
  \bibinfo {author} {\bibfnamefont {Cenke}\ \bibnamefont {Xu}}, \ and\ \bibinfo
  {author} {\bibfnamefont {T.}~\bibnamefont {Senthil}},\ }\bibfield  {title}
  {\enquote {\bibinfo {title} {Deconfined quantum critical points: Symmetries
  and dualities},}\ }\href {\doibase 10.1103/PhysRevX.7.031051} {\bibfield
  {journal} {\bibinfo  {journal} {Phys. Rev. X}\ }\textbf {\bibinfo {volume}
  {7}},\ \bibinfo {pages} {031051} (\bibinfo {year} {2017})}\BibitemShut
  {NoStop}%
\bibitem [{\citenamefont {Qin}\ \emph {et~al.}(2017)\citenamefont {Qin},
  \citenamefont {He}, \citenamefont {You}, \citenamefont {Lu}, \citenamefont
  {Sen}, \citenamefont {Sandvik}, \citenamefont {Xu},\ and\ \citenamefont
  {Meng}}]{Sandvik_PhysRevX.7.031052}%
  \BibitemOpen
  \bibfield  {author} {\bibinfo {author} {\bibfnamefont {Yan~Qi}\ \bibnamefont
  {Qin}}, \bibinfo {author} {\bibfnamefont {Yuan-Yao}\ \bibnamefont {He}},
  \bibinfo {author} {\bibfnamefont {Yi-Zhuang}\ \bibnamefont {You}}, \bibinfo
  {author} {\bibfnamefont {Zhong-Yi}\ \bibnamefont {Lu}}, \bibinfo {author}
  {\bibfnamefont {Arnab}\ \bibnamefont {Sen}}, \bibinfo {author} {\bibfnamefont
  {Anders~W.}\ \bibnamefont {Sandvik}}, \bibinfo {author} {\bibfnamefont
  {Cenke}\ \bibnamefont {Xu}}, \ and\ \bibinfo {author} {\bibfnamefont
  {Zi~Yang}\ \bibnamefont {Meng}},\ }\bibfield  {title} {\enquote {\bibinfo
  {title} {Duality between the deconfined quantum-critical point and the
  bosonic topological transition},}\ }\href {\doibase
  10.1103/PhysRevX.7.031052} {\bibfield  {journal} {\bibinfo  {journal} {Phys.
  Rev. X}\ }\textbf {\bibinfo {volume} {7}},\ \bibinfo {pages} {031052}
  (\bibinfo {year} {2017})}\BibitemShut {NoStop}%
\bibitem [{\citenamefont {Isakov}\ \emph {et~al.}(2012)\citenamefont {Isakov},
  \citenamefont {Melko},\ and\ \citenamefont {Hastings}}]{Isakov193}%
  \BibitemOpen
  \bibfield  {author} {\bibinfo {author} {\bibfnamefont {Sergei~V.}\
  \bibnamefont {Isakov}}, \bibinfo {author} {\bibfnamefont {Roger~G.}\
  \bibnamefont {Melko}}, \ and\ \bibinfo {author} {\bibfnamefont {Matthew~B.}\
  \bibnamefont {Hastings}},\ }\bibfield  {title} {\enquote {\bibinfo {title}
  {Universal signatures of fractionalized quantum critical points},}\ }\href
  {\doibase 10.1126/science.1212207} {\bibfield  {journal} {\bibinfo  {journal}
  {Science}\ }\textbf {\bibinfo {volume} {335}},\ \bibinfo {pages} {193--195}
  (\bibinfo {year} {2012})}\BibitemShut {NoStop}%
\bibitem [{\citenamefont {Kleinert}(1989)}]{kleinert1989gauge}%
  \BibitemOpen
  \bibfield  {author} {\bibinfo {author} {\bibfnamefont {Hagen}\ \bibnamefont
  {Kleinert}},\ }\href@noop {} {\emph {\bibinfo {title} {Gauge Fields in
  Condensed Matter: Vol. 1: Superflow and Vortex Lines (Disorder Fields, Phase
  Transitions) Vol. 2: Stresses and Defects (Differential Geometry, Crystal
  Melting)}}}\ (\bibinfo  {publisher} {World Scientific},\ \bibinfo {year}
  {1989})\BibitemShut {NoStop}%
\bibitem [{\citenamefont {Peskin}(1978)}]{Peskin1978}%
  \BibitemOpen
  \bibfield  {author} {\bibinfo {author} {\bibfnamefont {Michael~E.}\
  \bibnamefont {Peskin}},\ }\bibfield  {title} {\enquote {\bibinfo {title}
  {{Mandelstam-'t Hooft duality in abelian lattice models}},}\ }\href {\doibase
  10.1016/0003-4916(78)90252-X} {\bibfield  {journal} {\bibinfo  {journal}
  {Ann. Phys. (N. Y).}\ }\textbf {\bibinfo {volume} {113}},\ \bibinfo {pages}
  {122--152} (\bibinfo {year} {1978})}\BibitemShut {NoStop}%
\bibitem [{\citenamefont {Fr{\"o}hlich}\ and\ \citenamefont
  {Marchetti}(1989)}]{Frohlich-Marchetti}%
  \BibitemOpen
  \bibfield  {author} {\bibinfo {author} {\bibfnamefont {J.}~\bibnamefont
  {Fr{\"o}hlich}}\ and\ \bibinfo {author} {\bibfnamefont {P.~A.}\ \bibnamefont
  {Marchetti}},\ }\bibfield  {title} {\enquote {\bibinfo {title} {Quantum field
  theories of vortices and anyons},}\ }\href {\doibase 10.1007/BF01217803}
  {\bibfield  {journal} {\bibinfo  {journal} {Communications in Mathematical
  Physics}\ }\textbf {\bibinfo {volume} {121}},\ \bibinfo {pages} {177--223}
  (\bibinfo {year} {1989})}\BibitemShut {NoStop}%
\bibitem [{\citenamefont {Eliezer}\ and\ \citenamefont
  {Semenoff}(1992)}]{ELIEZER1992118}%
  \BibitemOpen
  \bibfield  {author} {\bibinfo {author} {\bibfnamefont {D.}~\bibnamefont
  {Eliezer}}\ and\ \bibinfo {author} {\bibfnamefont {G.W.}\ \bibnamefont
  {Semenoff}},\ }\bibfield  {title} {\enquote {\bibinfo {title} {Intersection
  forms and the geometry of lattice chern-simons theory},}\ }\href {\doibase
  https://doi.org/10.1016/0370-2693(92)90168-4} {\bibfield  {journal} {\bibinfo
   {journal} {Physics Letters B}\ }\textbf {\bibinfo {volume} {286}},\ \bibinfo
  {pages} {118 -- 124} (\bibinfo {year} {1992})}\BibitemShut {NoStop}%
\bibitem [{\citenamefont {Berruto}\ \emph {et~al.}(2000)\citenamefont
  {Berruto}, \citenamefont {Diamantini},\ and\ \citenamefont
  {Sodano}}]{BERRUTO2000366}%
  \BibitemOpen
  \bibfield  {author} {\bibinfo {author} {\bibfnamefont {F.}~\bibnamefont
  {Berruto}}, \bibinfo {author} {\bibfnamefont {M.C.}\ \bibnamefont
  {Diamantini}}, \ and\ \bibinfo {author} {\bibfnamefont {P.}~\bibnamefont
  {Sodano}},\ }\bibfield  {title} {\enquote {\bibinfo {title} {On pure lattice
  chern–simons gauge theories},}\ }\href {\doibase
  https://doi.org/10.1016/S0370-2693(00)00803-0} {\bibfield  {journal}
  {\bibinfo  {journal} {Physics Letters B}\ }\textbf {\bibinfo {volume}
  {487}},\ \bibinfo {pages} {366 -- 370} (\bibinfo {year} {2000})}\BibitemShut
  {NoStop}%
\bibitem [{\citenamefont {Gu}\ and\ \citenamefont
  {Wen}(2014)}]{Gu-Wen_PhysRevB.90.115141}%
  \BibitemOpen
  \bibfield  {author} {\bibinfo {author} {\bibfnamefont {Zheng-Cheng}\
  \bibnamefont {Gu}}\ and\ \bibinfo {author} {\bibfnamefont {Xiao-Gang}\
  \bibnamefont {Wen}},\ }\bibfield  {title} {\enquote {\bibinfo {title}
  {Symmetry-protected topological orders for interacting fermions: Fermionic
  topological nonlinear $\ensuremath{\sigma}$ models and a special group
  supercohomology theory},}\ }\href {\doibase 10.1103/PhysRevB.90.115141}
  {\bibfield  {journal} {\bibinfo  {journal} {Phys. Rev. B}\ }\textbf {\bibinfo
  {volume} {90}},\ \bibinfo {pages} {115141} (\bibinfo {year}
  {2014})}\BibitemShut {NoStop}%
\bibitem [{\citenamefont {Gaiotto}\ and\ \citenamefont
  {Kapustin}(2016)}]{Gaiotto-Kapustin}%
  \BibitemOpen
  \bibfield  {author} {\bibinfo {author} {\bibfnamefont {Davide}\ \bibnamefont
  {Gaiotto}}\ and\ \bibinfo {author} {\bibfnamefont {Anton}\ \bibnamefont
  {Kapustin}},\ }\bibfield  {title} {\enquote {\bibinfo {title} {Spin tqfts and
  fermionic phases of matter},}\ }\href {\doibase 10.1142/S0217751X16450445}
  {\bibfield  {journal} {\bibinfo  {journal} {International Journal of Modern
  Physics A}\ }\textbf {\bibinfo {volume} {31}},\ \bibinfo {pages} {1645044}
  (\bibinfo {year} {2016})}\BibitemShut {NoStop}%
\bibitem [{\citenamefont {DeMarco}\ and\ \citenamefont
  {Wen}(2021)}]{demarco2019lattice}%
  \BibitemOpen
  \bibfield  {author} {\bibinfo {author} {\bibfnamefont {Michael}\ \bibnamefont
  {DeMarco}}\ and\ \bibinfo {author} {\bibfnamefont {Xiao-Gang}\ \bibnamefont
  {Wen}},\ }\bibfield  {title} {\enquote {\bibinfo {title} {Compact
  ${\mathrm{u}}^{k}(1)$ chern-simons theory as a local bosonic lattice model
  with exact discrete 1-symmetries},}\ }\href {\doibase
  10.1103/PhysRevLett.126.021603} {\bibfield  {journal} {\bibinfo  {journal}
  {Phys. Rev. Lett.}\ }\textbf {\bibinfo {volume} {126}},\ \bibinfo {pages}
  {021603} (\bibinfo {year} {2021})}\BibitemShut {NoStop}%
\bibitem [{\citenamefont {Hansson}\ \emph {et~al.}(1989)\citenamefont
  {Hansson}, \citenamefont {Karlhede},\ and\ \citenamefont
  {Ro\v{c}ek}}]{HanssonKarlhedeRocek}%
  \BibitemOpen
  \bibfield  {author} {\bibinfo {author} {\bibfnamefont {T.H.}\ \bibnamefont
  {Hansson}}, \bibinfo {author} {\bibfnamefont {A.}~\bibnamefont {Karlhede}}, \
  and\ \bibinfo {author} {\bibfnamefont {M.}~\bibnamefont {Ro\v{c}ek}},\
  }\bibfield  {title} {\enquote {\bibinfo {title} {On wilson loops in abelian
  chern-simons theories},}\ }\href {\doibase
  https://doi.org/10.1016/0370-2693(89)91015-0} {\bibfield  {journal} {\bibinfo
   {journal} {Physics Letters B}\ }\textbf {\bibinfo {volume} {225}},\ \bibinfo
  {pages} {92--94} (\bibinfo {year} {1989})}\BibitemShut {NoStop}%
\bibitem [{\citenamefont {Thomas}\ and\ \citenamefont
  {Stone}(1978)}]{THOMAS1978513}%
  \BibitemOpen
  \bibfield  {author} {\bibinfo {author} {\bibfnamefont {Paul~R.}\ \bibnamefont
  {Thomas}}\ and\ \bibinfo {author} {\bibfnamefont {Michael}\ \bibnamefont
  {Stone}},\ }\bibfield  {title} {\enquote {\bibinfo {title} {Nature of the
  phase transition in a non-linear o(2)3 model},}\ }\href {\doibase
  https://doi.org/10.1016/0550-3213(78)90383-8} {\bibfield  {journal} {\bibinfo
   {journal} {Nuclear Physics B}\ }\textbf {\bibinfo {volume} {144}},\ \bibinfo
  {pages} {513 -- 524} (\bibinfo {year} {1978})}\BibitemShut {NoStop}%
\bibitem [{\citenamefont {Dasgupta}\ and\ \citenamefont
  {Halperin}(1981)}]{Dasgupta-Halperin_PhysRevLett.47.1556}%
  \BibitemOpen
  \bibfield  {author} {\bibinfo {author} {\bibfnamefont {C.}~\bibnamefont
  {Dasgupta}}\ and\ \bibinfo {author} {\bibfnamefont {B.~I.}\ \bibnamefont
  {Halperin}},\ }\bibfield  {title} {\enquote {\bibinfo {title} {Phase
  transition in a lattice model of superconductivity},}\ }\href {\doibase
  10.1103/PhysRevLett.47.1556} {\bibfield  {journal} {\bibinfo  {journal}
  {Phys. Rev. Lett.}\ }\textbf {\bibinfo {volume} {47}},\ \bibinfo {pages}
  {1556--1560} (\bibinfo {year} {1981})}\BibitemShut {NoStop}%
\bibitem [{\citenamefont {Polyakov}(1988)}]{Polyakov-1988}%
  \BibitemOpen
  \bibfield  {author} {\bibinfo {author} {\bibfnamefont {A.~M.}\ \bibnamefont
  {Polyakov}},\ }\bibfield  {title} {\enquote {\bibinfo {title} {Fermi-bose
  transmutations induced by gauge fields},}\ }\href {\doibase
  10.1142/S0217732388000398} {\bibfield  {journal} {\bibinfo  {journal} {Modern
  Physics Letters A}\ }\textbf {\bibinfo {volume} {03}},\ \bibinfo {pages}
  {325--328} (\bibinfo {year} {1988})}\BibitemShut {NoStop}%
\bibitem [{\citenamefont {T\"urker}\ \emph {et~al.}(2020)\citenamefont
  {T\"urker}, \citenamefont {van~den Brink}, \citenamefont {Meng},\ and\
  \citenamefont {Nogueira}}]{turker2020bosonization}%
  \BibitemOpen
  \bibfield  {author} {\bibinfo {author} {\bibfnamefont {Oguz}\ \bibnamefont
  {T\"urker}}, \bibinfo {author} {\bibfnamefont {Jeroen}\ \bibnamefont {van~den
  Brink}}, \bibinfo {author} {\bibfnamefont {Tobias}\ \bibnamefont {Meng}}, \
  and\ \bibinfo {author} {\bibfnamefont {Flavio~S.}\ \bibnamefont {Nogueira}},\
  }\bibfield  {title} {\enquote {\bibinfo {title} {Bosonization in $2+1$
  dimensions via chern-simons bosonic particle-vortex duality},}\ }\href
  {\doibase 10.1103/PhysRevD.102.034506} {\bibfield  {journal} {\bibinfo
  {journal} {Phys. Rev. D}\ }\textbf {\bibinfo {volume} {102}},\ \bibinfo
  {pages} {034506} (\bibinfo {year} {2020})}\BibitemShut {NoStop}%
\bibitem [{\citenamefont {Gau\ss}(1877)}]{cf5mathematischen}%
  \BibitemOpen
  \bibfield  {author} {\bibinfo {author} {\bibfnamefont {C~F}\ \bibnamefont
  {Gau\ss}},\ }\bibfield  {title} {\enquote {\bibinfo {title} {Zur
  mathematischen theorie der elektrodynamischen wirkungen},}\ }\href@noop {}
  {\bibfield  {journal} {\bibinfo  {journal} {Gau{\ss} CF Werke. Bd}\ }\textbf
  {\bibinfo {volume} {5}},\ \bibinfo {pages} {601--630} (\bibinfo {year}
  {1877})}\BibitemShut {NoStop}%
\bibitem [{\citenamefont {Frankel}(2011)}]{frankel2011geometry}%
  \BibitemOpen
  \bibfield  {author} {\bibinfo {author} {\bibfnamefont {Theodore}\
  \bibnamefont {Frankel}},\ }\href@noop {} {\emph {\bibinfo {title} {The
  geometry of physics: an introduction}}}\ (\bibinfo  {publisher} {Cambridge
  university press},\ \bibinfo {year} {2011})\BibitemShut {NoStop}%
\bibitem [{\citenamefont
  {C{\u{a}}lug{\u{a}}reanu}(1959)}]{calugareanu1959integrale}%
  \BibitemOpen
  \bibfield  {author} {\bibinfo {author} {\bibfnamefont {George}\ \bibnamefont
  {C{\u{a}}lug{\u{a}}reanu}},\ }\bibfield  {title} {\enquote {\bibinfo {title}
  {L’int{\'e}grale de gauss et l’analyse des n{\oe}uds tridimensionnels},}\
  }\href@noop {} {\bibfield  {journal} {\bibinfo  {journal} {Rev. Math. pures
  appl}\ }\textbf {\bibinfo {volume} {4}},\ \bibinfo {pages} {5--20} (\bibinfo
  {year} {1959})}\BibitemShut {NoStop}%
\bibitem [{\citenamefont
  {C{\u{a}}lug{\u{a}}reanu}(1961)}]{calugareanu1961classes}%
  \BibitemOpen
  \bibfield  {author} {\bibinfo {author} {\bibfnamefont {G}~\bibnamefont
  {C{\u{a}}lug{\u{a}}reanu}},\ }\bibfield  {title} {\enquote {\bibinfo {title}
  {Sur les classes d'isotopie des noeuds tridimensionnels et leurs
  invariants},}\ }\href@noop {} {\bibfield  {journal} {\bibinfo  {journal}
  {Czechoslovak Mathematical Journal}\ }\textbf {\bibinfo {volume} {11}},\
  \bibinfo {pages} {588--625} (\bibinfo {year} {1961})}\BibitemShut {NoStop}%
\bibitem [{\citenamefont {White}(1969)}]{White}%
  \BibitemOpen
  \bibfield  {author} {\bibinfo {author} {\bibfnamefont {James~H.}\
  \bibnamefont {White}},\ }\bibfield  {title} {\enquote {\bibinfo {title}
  {Self-linking and the gauss integral in higher dimensions},}\ }\href
  {http://www.jstor.org/stable/2373348} {\bibfield  {journal} {\bibinfo
  {journal} {American Journal of Mathematics}\ }\textbf {\bibinfo {volume}
  {91}},\ \bibinfo {pages} {693--728} (\bibinfo {year} {1969})}\BibitemShut
  {NoStop}%
\bibitem [{\citenamefont {Witten}(1989)}]{Witten_Jones}%
  \BibitemOpen
  \bibfield  {author} {\bibinfo {author} {\bibfnamefont {Edward}\ \bibnamefont
  {Witten}},\ }\bibfield  {title} {\enquote {\bibinfo {title} {{Quantum Field
  Theory and the Jones Polynomial}},}\ }\href {\doibase 10.1007/BF01217730}
  {\bibfield  {journal} {\bibinfo  {journal} {Commun. Math. Phys.}\ }\textbf
  {\bibinfo {volume} {121}},\ \bibinfo {pages} {351--399} (\bibinfo {year}
  {1989})}\BibitemShut {NoStop}%
\bibitem [{\citenamefont
  {Redlich}(1984{\natexlab{a}})}]{Redlich_PhysRevLett.52.18}%
  \BibitemOpen
  \bibfield  {author} {\bibinfo {author} {\bibfnamefont {A.~N.}\ \bibnamefont
  {Redlich}},\ }\bibfield  {title} {\enquote {\bibinfo {title} {Gauge
  noninvariance and parity nonconservation of three-dimensional fermions},}\
  }\href {\doibase 10.1103/PhysRevLett.52.18} {\bibfield  {journal} {\bibinfo
  {journal} {Phys. Rev. Lett.}\ }\textbf {\bibinfo {volume} {52}},\ \bibinfo
  {pages} {18--21} (\bibinfo {year} {1984}{\natexlab{a}})}\BibitemShut
  {NoStop}%
\bibitem [{\citenamefont
  {Redlich}(1984{\natexlab{b}})}]{Redlich_PhysRevD.29.2366}%
  \BibitemOpen
  \bibfield  {author} {\bibinfo {author} {\bibfnamefont {A.~N.}\ \bibnamefont
  {Redlich}},\ }\bibfield  {title} {\enquote {\bibinfo {title} {Parity
  violation and gauge noninvariance of the effective gauge field action in
  three dimensions},}\ }\href {\doibase 10.1103/PhysRevD.29.2366} {\bibfield
  {journal} {\bibinfo  {journal} {Phys. Rev. D}\ }\textbf {\bibinfo {volume}
  {29}},\ \bibinfo {pages} {2366--2374} (\bibinfo {year}
  {1984}{\natexlab{b}})}\BibitemShut {NoStop}%
\bibitem [{\citenamefont {Ferreiros}\ and\ \citenamefont
  {Fradkin}(2018)}]{FERREIROS20181}%
  \BibitemOpen
  \bibfield  {author} {\bibinfo {author} {\bibfnamefont {Yago}\ \bibnamefont
  {Ferreiros}}\ and\ \bibinfo {author} {\bibfnamefont {Eduardo}\ \bibnamefont
  {Fradkin}},\ }\bibfield  {title} {\enquote {\bibinfo {title} {Boson–fermion
  duality in a gravitational background},}\ }\href {\doibase
  https://doi.org/10.1016/j.aop.2018.10.001} {\bibfield  {journal} {\bibinfo
  {journal} {Annals of Physics}\ }\textbf {\bibinfo {volume} {399}},\ \bibinfo
  {pages} {1 -- 25} (\bibinfo {year} {2018})}\BibitemShut {NoStop}%
\bibitem [{\citenamefont {Polyakov}(1990)}]{polyakov1990two}%
  \BibitemOpen
  \bibfield  {author} {\bibinfo {author} {\bibfnamefont {AM}~\bibnamefont
  {Polyakov}},\ }\bibfield  {title} {\enquote {\bibinfo {title}
  {Two-dimensional quantum gravity: Superconductivity at high t\_c},}\
  }\href@noop {} {\bibfield  {journal} {\bibinfo  {journal} {Les Houches 1988,
  Proceedings, Fields, strings and critical phenomena}\ } (\bibinfo {year}
  {1990})}\BibitemShut {NoStop}%
\bibitem [{\citenamefont {Ambj{\o}rn}\ \emph {et~al.}(1990)\citenamefont
  {Ambj{\o}rn}, \citenamefont {Durhuus},\ and\ \citenamefont
  {Jonsson}}]{AMBJORN1990509}%
  \BibitemOpen
  \bibfield  {author} {\bibinfo {author} {\bibfnamefont {Jan}\ \bibnamefont
  {Ambj{\o}rn}}, \bibinfo {author} {\bibfnamefont {Bergfinnur}\ \bibnamefont
  {Durhuus}}, \ and\ \bibinfo {author} {\bibfnamefont {Thordur}\ \bibnamefont
  {Jonsson}},\ }\bibfield  {title} {\enquote {\bibinfo {title} {A random walk
  representation of the dirac propagator},}\ }\href {\doibase
  https://doi.org/10.1016/0550-3213(90)90121-S} {\bibfield  {journal} {\bibinfo
   {journal} {Nuclear Physics B}\ }\textbf {\bibinfo {volume} {330}},\ \bibinfo
  {pages} {509 -- 522} (\bibinfo {year} {1990})}\BibitemShut {NoStop}%
\bibitem [{\citenamefont {Grundberg}\ \emph {et~al.}(1990)\citenamefont
  {Grundberg}, \citenamefont {Hansson},\ and\ \citenamefont
  {Karlhede}}]{Grundberg1990}%
  \BibitemOpen
  \bibfield  {author} {\bibinfo {author} {\bibfnamefont {J.}~\bibnamefont
  {Grundberg}}, \bibinfo {author} {\bibfnamefont {T.H.}\ \bibnamefont
  {Hansson}}, \ and\ \bibinfo {author} {\bibfnamefont {A.}~\bibnamefont
  {Karlhede}},\ }\bibfield  {title} {\enquote {\bibinfo {title} {{On Polyakov's
  spin factors}},}\ }\href {\doibase 10.1016/0550-3213(90)90566-V} {\bibfield
  {journal} {\bibinfo  {journal} {Nucl. Phys. B}\ }\textbf {\bibinfo {volume}
  {347}},\ \bibinfo {pages} {420--440} (\bibinfo {year} {1990})}\BibitemShut
  {NoStop}%
\bibitem [{\citenamefont {Goldman}\ and\ \citenamefont
  {Fradkin}(2018)}]{Goldman2018}%
  \BibitemOpen
  \bibfield  {author} {\bibinfo {author} {\bibfnamefont {Hart}\ \bibnamefont
  {Goldman}}\ and\ \bibinfo {author} {\bibfnamefont {Eduardo}\ \bibnamefont
  {Fradkin}},\ }\bibfield  {title} {\enquote {\bibinfo {title} {Loop models,
  modular invariance, and three-dimensional bosonization},}\ }\href {\doibase
  10.1103/PhysRevB.97.195112} {\bibfield  {journal} {\bibinfo  {journal} {Phys.
  Rev. B}\ }\textbf {\bibinfo {volume} {97}},\ \bibinfo {pages} {195112}
  (\bibinfo {year} {2018})}\BibitemShut {NoStop}%
\bibitem [{\citenamefont {Witten}(1982)}]{WITTEN1982324}%
  \BibitemOpen
  \bibfield  {author} {\bibinfo {author} {\bibfnamefont {Edward}\ \bibnamefont
  {Witten}},\ }\bibfield  {title} {\enquote {\bibinfo {title} {An su(2)
  anomaly},}\ }\href {\doibase https://doi.org/10.1016/0370-2693(82)90728-6}
  {\bibfield  {journal} {\bibinfo  {journal} {Physics Letters B}\ }\textbf
  {\bibinfo {volume} {117}},\ \bibinfo {pages} {324 -- 328} (\bibinfo {year}
  {1982})}\BibitemShut {NoStop}%
\bibitem [{\citenamefont {Alvarez-Gaumé}\ \emph {et~al.}(1985)\citenamefont
  {Alvarez-Gaumé}, \citenamefont {{Della Pietra}},\ and\ \citenamefont
  {Moore}}]{ALVAREZGAUME1985288}%
  \BibitemOpen
  \bibfield  {author} {\bibinfo {author} {\bibfnamefont {L}~\bibnamefont
  {Alvarez-Gaumé}}, \bibinfo {author} {\bibfnamefont {S}~\bibnamefont {{Della
  Pietra}}}, \ and\ \bibinfo {author} {\bibfnamefont {G}~\bibnamefont
  {Moore}},\ }\bibfield  {title} {\enquote {\bibinfo {title} {Anomalies and odd
  dimensions},}\ }\href {\doibase https://doi.org/10.1016/0003-4916(85)90383-5}
  {\bibfield  {journal} {\bibinfo  {journal} {Annals of Physics}\ }\textbf
  {\bibinfo {volume} {163}},\ \bibinfo {pages} {288 -- 317} (\bibinfo {year}
  {1985})}\BibitemShut {NoStop}%
\bibitem [{\citenamefont {Forte}(1987)}]{FORTE1987252}%
  \BibitemOpen
  \bibfield  {author} {\bibinfo {author} {\bibfnamefont {Stefano}\ \bibnamefont
  {Forte}},\ }\bibfield  {title} {\enquote {\bibinfo {title} {Explicit
  construction of anomalies},}\ }\href {\doibase
  https://doi.org/10.1016/0550-3213(87)90215-X} {\bibfield  {journal} {\bibinfo
   {journal} {Nuclear Physics B}\ }\textbf {\bibinfo {volume} {288}},\ \bibinfo
  {pages} {252 -- 274} (\bibinfo {year} {1987})}\BibitemShut {NoStop}%
\bibitem [{\citenamefont {Chen}\ \emph {et~al.}(2018)\citenamefont {Chen},
  \citenamefont {Son}, \citenamefont {Wang},\ and\ \citenamefont
  {Raghu}}]{Raghu_PhysRevLett.120.016602}%
  \BibitemOpen
  \bibfield  {author} {\bibinfo {author} {\bibfnamefont {Jing-Yuan}\
  \bibnamefont {Chen}}, \bibinfo {author} {\bibfnamefont {Jun~Ho}\ \bibnamefont
  {Son}}, \bibinfo {author} {\bibfnamefont {Chao}\ \bibnamefont {Wang}}, \ and\
  \bibinfo {author} {\bibfnamefont {S.}~\bibnamefont {Raghu}},\ }\bibfield
  {title} {\enquote {\bibinfo {title} {Exact boson-fermion duality on a 3d
  euclidean lattice},}\ }\href {\doibase 10.1103/PhysRevLett.120.016602}
  {\bibfield  {journal} {\bibinfo  {journal} {Phys. Rev. Lett.}\ }\textbf
  {\bibinfo {volume} {120}},\ \bibinfo {pages} {016602} (\bibinfo {year}
  {2018})}\BibitemShut {NoStop}%
\bibitem [{\citenamefont {Aharony}(2016)}]{Aharony-JHEP-2016}%
  \BibitemOpen
  \bibfield  {author} {\bibinfo {author} {\bibfnamefont {Ofer}\ \bibnamefont
  {Aharony}},\ }\bibfield  {title} {\enquote {\bibinfo {title} {Baryons,
  monopoles and dualities in chern-simons-matter theories},}\ }\href {\doibase
  10.1007/JHEP02(2016)093} {\bibfield  {journal} {\bibinfo  {journal} {Journal
  of High Energy Physics}\ }\textbf {\bibinfo {volume} {2016}},\ \bibinfo
  {pages} {93} (\bibinfo {year} {2016})}\BibitemShut {NoStop}%
\bibitem [{\citenamefont {Hsin}\ and\ \citenamefont
  {Seiberg}(2016)}]{Hsin-Seiberg}%
  \BibitemOpen
  \bibfield  {author} {\bibinfo {author} {\bibfnamefont {Po-Shen}\ \bibnamefont
  {Hsin}}\ and\ \bibinfo {author} {\bibfnamefont {Nathan}\ \bibnamefont
  {Seiberg}},\ }\bibfield  {title} {\enquote {\bibinfo {title} {Level/rank
  duality and chern-simons-matter theories},}\ }\href {\doibase
  10.1007/JHEP09(2016)095} {\bibfield  {journal} {\bibinfo  {journal} {Journal
  of High Energy Physics}\ }\textbf {\bibinfo {volume} {2016}},\ \bibinfo
  {pages} {95} (\bibinfo {year} {2016})}\BibitemShut {NoStop}%
\bibitem [{\citenamefont {Wen}\ and\ \citenamefont
  {Zee}(1992)}]{Wen-Zee_PhysRevLett.69.1811}%
  \BibitemOpen
  \bibfield  {author} {\bibinfo {author} {\bibfnamefont {Xiao-Gang}\
  \bibnamefont {Wen}}\ and\ \bibinfo {author} {\bibfnamefont {A.}~\bibnamefont
  {Zee}},\ }\bibfield  {title} {\enquote {\bibinfo {title} {Neutral superfluid
  modes and ``magnetic'' monopoles in multilayered quantum hall systems},}\
  }\href {\doibase 10.1103/PhysRevLett.69.1811} {\bibfield  {journal} {\bibinfo
   {journal} {Phys. Rev. Lett.}\ }\textbf {\bibinfo {volume} {69}},\ \bibinfo
  {pages} {1811--1814} (\bibinfo {year} {1992})}\BibitemShut {NoStop}%
\bibitem [{\citenamefont {Wen}(2004)}]{wen2004quantum}%
  \BibitemOpen
  \bibfield  {author} {\bibinfo {author} {\bibfnamefont {Xiao-Gang}\
  \bibnamefont {Wen}},\ }\href@noop {} {\emph {\bibinfo {title} {Quantum field
  theory of many-body systems: from the origin of sound to an origin of light
  and electrons}}}\ (\bibinfo  {publisher} {Oxford University Press on
  Demand},\ \bibinfo {year} {2004})\BibitemShut {NoStop}%
\bibitem [{\citenamefont {Kim}\ \emph {et~al.}(2001)\citenamefont {Kim},
  \citenamefont {Nayak}, \citenamefont {Demler}, \citenamefont {Read},\ and\
  \citenamefont {Das~Sarma}}]{Kim_PhysRevB.63.205315}%
  \BibitemOpen
  \bibfield  {author} {\bibinfo {author} {\bibfnamefont {Yong~Baek}\
  \bibnamefont {Kim}}, \bibinfo {author} {\bibfnamefont {Chetan}\ \bibnamefont
  {Nayak}}, \bibinfo {author} {\bibfnamefont {Eugene}\ \bibnamefont {Demler}},
  \bibinfo {author} {\bibfnamefont {N.}~\bibnamefont {Read}}, \ and\ \bibinfo
  {author} {\bibfnamefont {S.}~\bibnamefont {Das~Sarma}},\ }\bibfield  {title}
  {\enquote {\bibinfo {title} {Bilayer paired quantum hall states and coulomb
  drag},}\ }\href {\doibase 10.1103/PhysRevB.63.205315} {\bibfield  {journal}
  {\bibinfo  {journal} {Phys. Rev. B}\ }\textbf {\bibinfo {volume} {63}},\
  \bibinfo {pages} {205315} (\bibinfo {year} {2001})}\BibitemShut {NoStop}%
\bibitem [{\citenamefont {Spielman}\ \emph {et~al.}(2000)\citenamefont
  {Spielman}, \citenamefont {Eisenstein}, \citenamefont {Pfeiffer},\ and\
  \citenamefont {West}}]{Spielman_PhysRevLett.84.5808}%
  \BibitemOpen
  \bibfield  {author} {\bibinfo {author} {\bibfnamefont {I.~B.}\ \bibnamefont
  {Spielman}}, \bibinfo {author} {\bibfnamefont {J.~P.}\ \bibnamefont
  {Eisenstein}}, \bibinfo {author} {\bibfnamefont {L.~N.}\ \bibnamefont
  {Pfeiffer}}, \ and\ \bibinfo {author} {\bibfnamefont {K.~W.}\ \bibnamefont
  {West}},\ }\bibfield  {title} {\enquote {\bibinfo {title} {Resonantly
  enhanced tunneling in a double layer quantum hall ferromagnet},}\ }\href
  {\doibase 10.1103/PhysRevLett.84.5808} {\bibfield  {journal} {\bibinfo
  {journal} {Phys. Rev. Lett.}\ }\textbf {\bibinfo {volume} {84}},\ \bibinfo
  {pages} {5808--5811} (\bibinfo {year} {2000})}\BibitemShut {NoStop}%
\bibitem [{\citenamefont {Fogler}\ and\ \citenamefont
  {Wilczek}(2001)}]{Wilczek_PhysRevLett.86.1833}%
  \BibitemOpen
  \bibfield  {author} {\bibinfo {author} {\bibfnamefont {Michael~M.}\
  \bibnamefont {Fogler}}\ and\ \bibinfo {author} {\bibfnamefont {Frank}\
  \bibnamefont {Wilczek}},\ }\bibfield  {title} {\enquote {\bibinfo {title}
  {Josephson effect without superconductivity: Realization in quantum hall
  bilayers},}\ }\href {\doibase 10.1103/PhysRevLett.86.1833} {\bibfield
  {journal} {\bibinfo  {journal} {Phys. Rev. Lett.}\ }\textbf {\bibinfo
  {volume} {86}},\ \bibinfo {pages} {1833--1836} (\bibinfo {year}
  {2001})}\BibitemShut {NoStop}%
\bibitem [{\citenamefont {Balents}\ and\ \citenamefont
  {Radzihovsky}(2001)}]{Balents_PhysRevLett.86.1825}%
  \BibitemOpen
  \bibfield  {author} {\bibinfo {author} {\bibfnamefont {L.}~\bibnamefont
  {Balents}}\ and\ \bibinfo {author} {\bibfnamefont {L.}~\bibnamefont
  {Radzihovsky}},\ }\bibfield  {title} {\enquote {\bibinfo {title} {Interlayer
  tunneling in double-layer quantum hall pseudoferromagnets},}\ }\href
  {\doibase 10.1103/PhysRevLett.86.1825} {\bibfield  {journal} {\bibinfo
  {journal} {Phys. Rev. Lett.}\ }\textbf {\bibinfo {volume} {86}},\ \bibinfo
  {pages} {1825--1828} (\bibinfo {year} {2001})}\BibitemShut {NoStop}%
\bibitem [{\citenamefont {Stern}\ \emph {et~al.}(2001)\citenamefont {Stern},
  \citenamefont {Girvin}, \citenamefont {MacDonald},\ and\ \citenamefont
  {Ma}}]{Stern_PhysRevLett.86.1829}%
  \BibitemOpen
  \bibfield  {author} {\bibinfo {author} {\bibfnamefont {Ady}\ \bibnamefont
  {Stern}}, \bibinfo {author} {\bibfnamefont {S.~M.}\ \bibnamefont {Girvin}},
  \bibinfo {author} {\bibfnamefont {A.~H.}\ \bibnamefont {MacDonald}}, \ and\
  \bibinfo {author} {\bibfnamefont {Ning}\ \bibnamefont {Ma}},\ }\bibfield
  {title} {\enquote {\bibinfo {title} {Theory of interlayer tunneling in
  bilayer quantum hall ferromagnets},}\ }\href {\doibase
  10.1103/PhysRevLett.86.1829} {\bibfield  {journal} {\bibinfo  {journal}
  {Phys. Rev. Lett.}\ }\textbf {\bibinfo {volume} {86}},\ \bibinfo {pages}
  {1829--1832} (\bibinfo {year} {2001})}\BibitemShut {NoStop}%
\bibitem [{\citenamefont {Khalaf}\ \emph {et~al.}(2021)\citenamefont {Khalaf},
  \citenamefont {Chatterjee}, \citenamefont {Bultinck}, \citenamefont
  {Zaletel},\ and\ \citenamefont {Vishwanath}}]{Vishwanath-tbg}%
  \BibitemOpen
  \bibfield  {author} {\bibinfo {author} {\bibfnamefont {Eslam}\ \bibnamefont
  {Khalaf}}, \bibinfo {author} {\bibfnamefont {Shubhayu}\ \bibnamefont
  {Chatterjee}}, \bibinfo {author} {\bibfnamefont {Nick}\ \bibnamefont
  {Bultinck}}, \bibinfo {author} {\bibfnamefont {Michael~P.}\ \bibnamefont
  {Zaletel}}, \ and\ \bibinfo {author} {\bibfnamefont {Ashvin}\ \bibnamefont
  {Vishwanath}},\ }\bibfield  {title} {\enquote {\bibinfo {title} {Charged
  skyrmions and topological origin of superconductivity in magic-angle
  graphene},}\ }\href {\doibase 10.1126/sciadv.abf5299} {\bibfield  {journal}
  {\bibinfo  {journal} {Science Advances}\ }\textbf {\bibinfo {volume} {7}}
  (\bibinfo {year} {2021}),\ 10.1126/sciadv.abf5299}\BibitemShut {NoStop}%
\bibitem [{\citenamefont {Parisi}(1980)}]{Parisi}%
  \BibitemOpen
  \bibfield  {author} {\bibinfo {author} {\bibfnamefont {Giorgio}\ \bibnamefont
  {Parisi}},\ }\bibfield  {title} {\enquote {\bibinfo {title} {Field-theoretic
  approach to second-order phase transitions in two- and three-dimensional
  systems},}\ }\href {\doibase 10.1007/BF01014429} {\bibfield  {journal}
  {\bibinfo  {journal} {Journal of Statistical Physics}\ }\textbf {\bibinfo
  {volume} {23}},\ \bibinfo {pages} {49--82} (\bibinfo {year}
  {1980})}\BibitemShut {NoStop}%
\bibitem [{\citenamefont {Nogueira}\ \emph {et~al.}(2019)\citenamefont
  {Nogueira}, \citenamefont {van~den Brink},\ and\ \citenamefont
  {Sudb\o{}}}]{Conformality_loss}%
  \BibitemOpen
  \bibfield  {author} {\bibinfo {author} {\bibfnamefont {Flavio~S.}\
  \bibnamefont {Nogueira}}, \bibinfo {author} {\bibfnamefont {Jeroen}\
  \bibnamefont {van~den Brink}}, \ and\ \bibinfo {author} {\bibfnamefont
  {Asle}\ \bibnamefont {Sudb\o{}}},\ }\bibfield  {title} {\enquote {\bibinfo
  {title} {Conformality loss and quantum criticality in topological higgs
  electrodynamics in $2+1$ dimensions},}\ }\href {\doibase
  10.1103/PhysRevD.100.085005} {\bibfield  {journal} {\bibinfo  {journal}
  {Phys. Rev. D}\ }\textbf {\bibinfo {volume} {100}},\ \bibinfo {pages}
  {085005} (\bibinfo {year} {2019})}\BibitemShut {NoStop}%
\bibitem [{\citenamefont {Coleman}\ and\ \citenamefont
  {Hill}(1985)}]{COLEMAN1985184}%
  \BibitemOpen
  \bibfield  {author} {\bibinfo {author} {\bibfnamefont {Sidney}\ \bibnamefont
  {Coleman}}\ and\ \bibinfo {author} {\bibfnamefont {Brian}\ \bibnamefont
  {Hill}},\ }\bibfield  {title} {\enquote {\bibinfo {title} {No more
  corrections to the topological mass term in qed3},}\ }\href {\doibase
  https://doi.org/10.1016/0370-2693(85)90883-4} {\bibfield  {journal} {\bibinfo
   {journal} {Physics Letters B}\ }\textbf {\bibinfo {volume} {159}},\ \bibinfo
  {pages} {184 -- 188} (\bibinfo {year} {1985})}\BibitemShut {NoStop}%
\bibitem [{\citenamefont {Zinn-Justin}(2002)}]{zinn2002quantum}%
  \BibitemOpen
  \bibfield  {author} {\bibinfo {author} {\bibfnamefont {Jean}\ \bibnamefont
  {Zinn-Justin}},\ }\href@noop {} {\emph {\bibinfo {title} {Quantum field
  theory and critical phenomena}}},\ \bibinfo {edition} {4th}\ ed.\ (\bibinfo
  {publisher} {Clarendon Press},\ \bibinfo {year} {2002})\BibitemShut {NoStop}%
\end{thebibliography}%

\section*{Supplemental Material}

\section{Renormalization group analysis}
First we consider the easy-plane CP$^1$ model with a soft constraint and without any additional gauge terms,
\begin{eqnarray}
	\label{eq:CP1soft}
	\mathcal{L}&=&\sum_{a=1,2}|(\partial_\mu-ia_\mu)z_a|^2+	\frac{K}{2}(|z_1|^2- |z_2|^2)^2\nonumber\\
	&+&m_0^2(|z_1|^2+|z_2|^2)+\frac{u}{2}(|z_1|^2+|z_2|^2)^2.
\end{eqnarray}
%
%


We define the couplings $U=u+K$ and $V=u-K$ and derive  the renormalized coupings at one-loop order, 
\begin{equation}
	U_r=U-(5U^2+V^2)I(m^2),
\end{equation}
\begin{equation}
	V_r=V-2(V^2+2UV)I(m^2),
\end{equation}
where, 
\begin{equation}
	I(m^2)=\int_p\frac{1}{(p^2+m^2)^2}=\frac{m^{d-4}}{(4\pi)^{d/2}}\Gamma\left(2-
	\frac{d}{2}\right),
\end{equation}
with $m$ being the renormalized mass, and we have introduced the notation $\int_{p} = \int \frac{d^3p}{(2 \pi)^3}$. 
Since the CS is absent, we are generalizing the 
calculation to $d$ dimensions. 
We define the dimensionless renormalized couplings by $\hat{U}=m^{d-4}U_r$ and 
$\hat{V}=m^{d-4}V_r$, along with the rescalings $\hat{U}\to\hat{U}/c_d$, 
$\hat{V}\to\hat{V}/c_d$, where $c_d=(4-d)(4\pi)^{-d/2}\Gamma(2-d/2)$, so that the 
RG $\beta$ functions, $\beta_{\hat{U}}=md\hat{U}/dm$ and $\beta_{\hat{V}}=md\hat{V}/dm$ 
are obtained, 
\begin{equation}
	\label{Eq:beta-U}
	\beta_{\hat{U}}=m\frac{d\hat{U}}{dm}=-(4-d)\hat{U}+5\hat{U}^2+\hat{V}^2,
\end{equation}
\begin{equation}
	\label{Eq:beta-V}
	\beta_{\hat{V}}=m\frac{d\hat{V}}{dm}=-(4-d)\hat{V}+2(\hat{V}^2+2\hat{U}\hat{V}).
\end{equation}
We obtain three fixed points, namely, the Gaussian fixed point, $(\hat{U}_G,\hat{V}_G)=(0,0)$, 
the $XY$ fixed point, $(\hat{U}_{\rm XY},\hat{V}_{\rm XY})=(\epsilon/5,0)$, and the 
easy-plane anisotropy fixed point, $(\hat{U}_*,\hat{V}_*)=(\epsilon/6)(1,1)$, 
where $\epsilon=4-d$. The 
$XY$ fixed point is stable for $\hat{V}=0$, but becomes unstable for any small $\hat{V}$. 
Note that this fixed point actually corresponds $\hat{U}=2\hat{K}$, so this yields 
a fixed point $\hat{K}_*=\epsilon/10$ associated to the $O(2)\times O(2)$ symmetry.  
The fixed point $(\hat{U}_*,\hat{V}_*)$ 
actually corresponds to vanishing anisotropy, i.e., $\hat{K}=0$, since $\hat{U}_*=\hat{V}_*$, 
implying $\hat{u}_*=\epsilon/6$. Hence, this fixed point governs the 
$O(4)$ universality class. This fixed point becomes IR unstable only in a region 
where $K<0$, which would correspond to easy-axis rather than easy-plane anisotropy.

Next, we consider the coupling to the gauge field $a_\mu$. We include CS and 
Maxwell terms,  
\begin{equation}
	\mathcal{L}_{\rm gauge}=\frac{1}{2e^2}(\epsilon_{\mu\nu\lambda}\partial_\nu a_\lambda)^2+i\frac{\kappa}{2}\epsilon_{\mu\nu\lambda}a_\mu\partial_\nu a_\lambda.
\end{equation}
The gauge field propagator in the absence of interactions is given in the Landau gauge by, 
\begin{equation}
	D_{\mu\nu}(p)=\frac{1}{p^2+M^2}\left(
	\delta_{\mu\nu}-\frac{p_\mu p_\nu}{p^2}-M\epsilon_{\mu\nu\lambda}\frac{p_\lambda}{p^2}\right),
\end{equation}
where $M=e^2\kappa$, and where we used the rescaling $a_\mu\to ea_\mu$. 

Since the CS term is defined in three spacetime dimensions, the renormalized couplings $U_r$ and $V_r$ will be calculated for fixed dimensionality $d=3$ \cite{Parisi}. This has the drawback of making $\epsilon=4-d$ as a control parameter in principle  unavailable to us (see section \ref{sec:control} for a more thorough discussion on this point). Instead, we generalize the easy-plane model to a system featuring a global $O(N)\times O(N)$ symmetry with $N$ even and explicitly consider the large $N$ limit. This is achieved by considering $N/2$ complex fields $z_{1a}$ and $z_{2a}$ ($a=1,\dots,N/2$). In this case the special case where the anisotropy is absent (i.e., $K=0$) will correspond to an $O(2N)$ global symmetry.

\begin{figure}[h]
	\includegraphics[width=4cm]{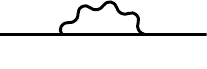}
	\caption{Scalar field self-energy}
	\label{Fig:self-energy}
\end{figure}

The coupling to a dynamical gauge field will cause $U_r$ and $V_r$ to   
receive a contribution from the 
diagram at Fig. \ref{Fig:self-energy} through the square of the wave function 
renormalization. This diagram is the only one giving a momentum dependent contribution to the total self-energy at one-loop. Its explicit expression is given by,  
\begin{eqnarray}
	\Sigma(p)&=&-e^2\int\frac{d^3k}{(2\pi)^3}\frac{(2p_\mu-k_\mu)(2p_\nu-k_\nu)}{(p-k)^2+m^2}D_{\mu\nu}(k) 
	\nonumber\\
	&=&4e^2\left\{\int\frac{d^3k}{(2\pi)^3}\frac{(k\cdot p)^2}{k^2[(p-k)^2+m^2](k^2+M^2)}
	\right.\nonumber\\
	&-&\left.\int\frac{d^3k}{(2\pi)^3}\frac{1}{[(p-k)^2+m^2](k^2+M^2)}	
	\right\}.
\end{eqnarray}
In the small external momentum $|p|$ limit, we obtain,  
\begin{equation}
	\Sigma(p)=-\frac{2e^2}{3\pi}\frac{p^2}{|m|+|M|}+\mathcal{O}(p^4).
\end{equation}

Another diagram contributing to both 
$U_r$ and $V_r$ is shown in Fig. \ref{Fig:photon4p-loop}. However, the 
latter vanishes at zero external momenta (see the Appendix A in Ref. \cite{Conformality_loss}). 

\begin{figure}[h]
	\includegraphics[width=4cm]{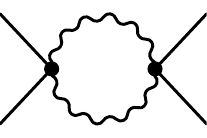}
	\caption{One-loop photon diagram contributing to the couplings $U_r$ and $V_r$.}
	\label{Fig:photon4p-loop}
\end{figure}

\begin{figure}[h]
	\includegraphics[width=4cm]{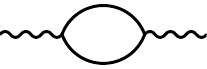}
	\caption{Vacuum polarization diagram. The wiggles represent the gauge field $a_\mu$. The continuous line can be either fermion or 
		boson propagators.}
	\label{Fig:poldiag}
\end{figure}

Therefore, we obtain the wave function renormalization, 
\begin{equation}
	Z=1+\frac{2}{3\pi}\frac{e^2}{m+e^2|\kappa|}
\end{equation}
and dimensionless renormalized couplings, 
\begin{eqnarray}
	\label{Eq:U-exp}
	\hat{U}&=&Z^2\left[\frac{U}{m}-\frac{(N+8)U^2+NV^2}{16\pi m^2}\right]
	\nonumber\\
	&\approx&\left[\frac{U}{m}-\frac{e^2}{2\pi m|\kappa|}+\frac{4e^2U}{3\pi m(m+e^2|\kappa|)}
	\right.\nonumber\\
	&-&\left.\frac{(N+8)U^2+NV^2}{16\pi m^2}\right],
\end{eqnarray}
\begin{eqnarray}
	\label{Eq:V-exp}
	\hat{V}&=&Z^2\left[\frac{V}{m}-\frac{2V^2+(N+2)UV}{8\pi m^2}\right]
	\nonumber\\
	&\approx&\left[\frac{V}{m}-\frac{e^2}{2\pi m|\kappa|}+\frac{4e^2V}{3\pi m(m+e^2|\kappa|)}
	\right.\nonumber\\
	&-&\left.\frac{2V^2+(N+2)UV}{8\pi m^2}\right],
\end{eqnarray}
where now the dimensionality is fixed to $d=3$. 
We define an additional dimensionless coupling, $\hat{e}^2=e_r^2/m$, where $e_r^2$ is the 
renormalized gauge coupling which is calculated at one-loop order by considering the vacuum polarization diagram of Fig. \ref{Fig:poldiag}. 

The new RG $\beta$ functions are given by, 
\begin{equation}
	\label{Eq:beta-U-1}
	\beta_{\hat{U}}=-\left[1+\frac{4\hat{e}^2}{3\pi(1+\hat{e}^2|\kappa|)^2}\right]\hat{U}+
	\frac{(N+8)\hat{U}^2+N\hat{V}^2}{16\pi},
\end{equation}
\begin{equation}
	\label{Eq:beta-V-1}
	\beta_{\hat{V}}=-\left[1+\frac{4\hat{e}^2}{3\pi(1+\hat{e}^2|\kappa|)^2}-
	\frac{2\hat{V}+(N+2)\hat{U}}{8\pi}\right]\hat{V},
\end{equation}
along with the $\beta$ function, 
\begin{equation}
	\label{Eq:betae2}
	\beta_{\hat{e}^2}=-\hat{e}^2+\frac{N\hat{e}^4}{24\pi}.
\end{equation}
On the other hand, the non-renormalization of the CS term \cite{COLEMAN1985184} implies, 
\begin{equation}
	\beta_\kappa=\left(\frac{N\hat{e}^2}{24\pi}-1\right)\kappa.
\end{equation}
Therefore, at the charged fixed point an arbitrary value of 
$\kappa$ is allowed, leading to a critical behavior featuring continuously varying 
critical exponents as a function of $\kappa$. The vanishing of $\beta_{\hat{e}^2}$ at the IR stable fixed point 
(i.e., $\hat{e}^2_*\neq 0$), automatically implies the vanishing of $\beta_\kappa$ for 
arbitrary $\kappa$. Thus, 
the fixed point structure of the $\beta$ functions (\ref{Eq:beta-U-1}) and 
(\ref{Eq:beta-V-1}) at the IR stable fixed point $\hat{e}^2_*=24\pi/N$ is similar to the one 
of the $\beta$ functions for the charge neutral system given by Eqs. (\ref{Eq:beta-U}) and 
(\ref{Eq:beta-V}). Plugging the fixed point $e_*^2$ into Eqs. (\ref{Eq:beta-U-1}) and (\ref{Eq:beta-V-1}) we find that these $\beta$ functions have two nontrivial fixed points, namely, 
\begin{equation}
	(\hat{U}_*,\hat{V}_*)=\left(\frac{16\pi[32N+(N+12n)^2]}{(N+8)(N+12n)^2},0\right), 
\end{equation}
corresponding to the $O(N)\times O(N)$ symmetry regime, 
while the $O(2N)$ symmetric case, 
\begin{equation}
	\hat{U}_*=\hat{V}_*=\frac{8\pi[32N+(N+12n)^2]}{(N+4)(N+12n)^2},
\end{equation}
where we have assumed a level $n$ CS term, $\kappa=n/(2\pi)$. 
An additional fixed point $(\hat{U}_1,\hat{V}_1)$ corresponding to a regime where $K>u$ is obtained for $N>4$ (recall that $N$ is even), where, 
\begin{equation}
	\hat{U}_1=\frac{8\pi[32N+(N+12n)^2]}{(N^2+8)(N+12n)^2},
\end{equation}
\begin{equation}
	\hat{V}_1=(4-N)\hat{U}_1.
\end{equation} 
Note that for $N=2$ the fixed point $(\hat{U}_1,\hat{V}_1)$ coincides with the $O(4)$ symmetric one.

\begin{figure}[t]
	\begin{center}
		\includegraphics[width=9cm]{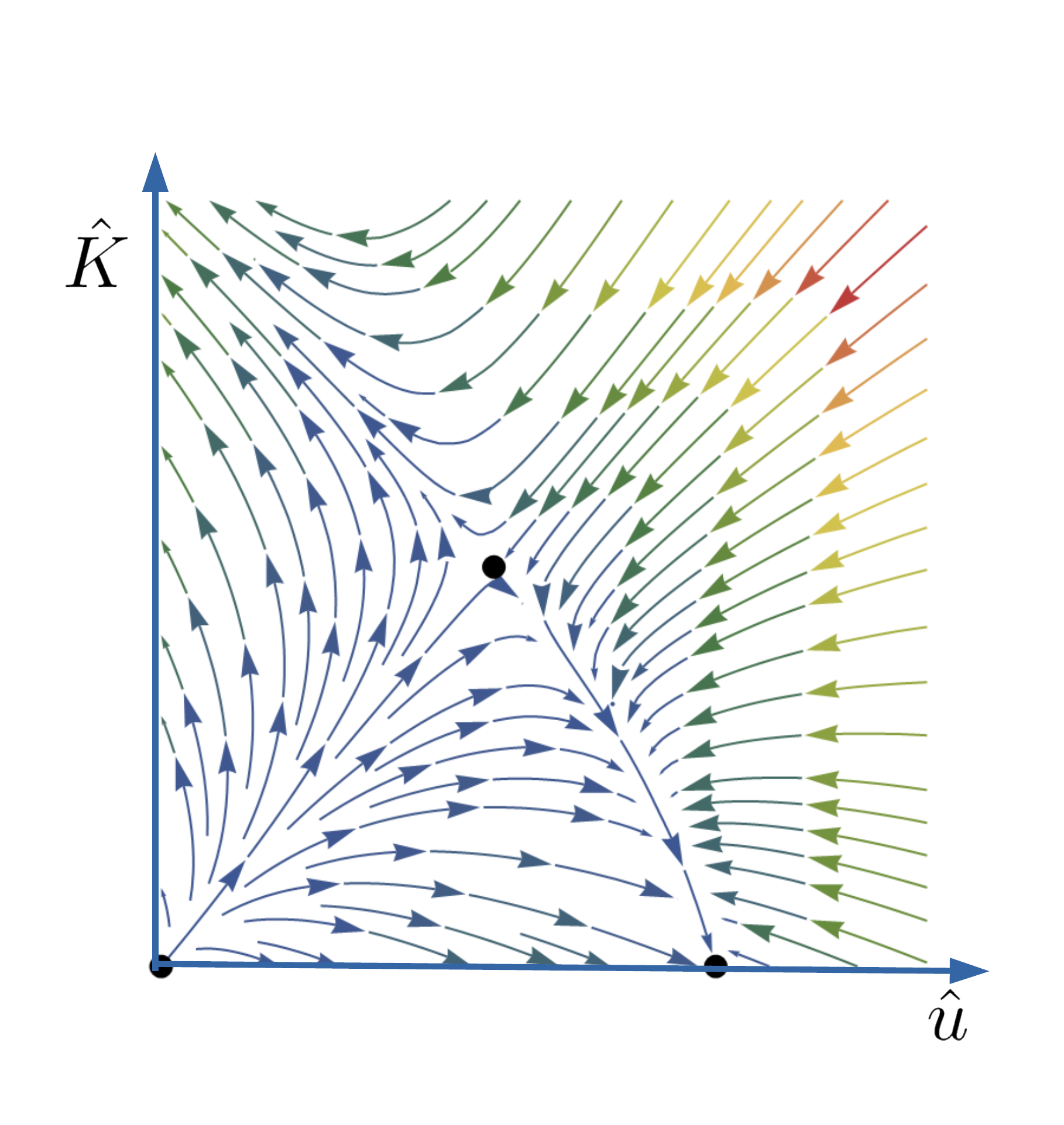}
		\caption{The RG flow for the $\beta$ functions of the dimensionless couplings $\hat{u}$ and $\hat{K}$. The $O(4)$ symmetric fixed point corrsponding to vanishing anisotropy is IR stable, while the anisotropy fixed point corresponding to an $O(2)\times O(2)$ symmetry is stable along the line $\hat{u}=\hat{K}$.}
		\label{Fig:RG-flow}
	\end{center}
\end{figure}

The flow diagram in terms of the original couplings $u$ and $K$ is shown in Fig. \ref{Fig:RG-flow}. Interestingly, we see that the $O(4)$ symmetric fixed point occurring for a vanishing anisotropy is IR stable. This implies that for the CS CP$^1$ theory a deconfined critical point occurs in the more symmetric case. The anisotropy fixed point is stable along the line $\hat{u}=\hat{K}$, corresponding to the case of a scalar self-coupling interaction of the form, $|z_1|^4+|z_2|^4$.

Hence, we arrive at two non-trivial fixed points that govern second-order 
phase transitions. Clearly, a new universality 
class emerges, since the critical exponents will depend on the CS level. 
For instance, for either the $O(2)\times O(2)$ or $O(4)$ 
symmetric fixed points, we obtain, 
\begin{equation}
	\frac{1}{\nu}=2-\frac{(\mathcal{N}+2)\hat{U}_*}{16\pi}+\frac{2\hat{e}^2_*}{3\pi(1+\hat{e}_*^2|\kappa|)^2},
\end{equation}
where $\mathcal{N}=2$ and $\mathcal{N}=4$ correspond to the $O(2)\times O(2)$ and $O(4)$ symmetries, respectively. 

Hence,   
\begin{equation}
	\frac{1}{\nu^{O(2)\times O(2)}}=\frac{8}{5}\left[1+\frac{1}{(1+6n)^2}\right],
\end{equation}
\begin{equation}
	\frac{1}{\nu^{O(4)}}=\frac{3}{2}. 
\end{equation}
Thus, for a level 1 CS term this yields $\nu^{O(2)\times O(2)}=49/80\approx 0.613$. This is nearly the same as 
the one-loop value $\nu=5/8$ of the $XY$ universality class. 

For the $O(4)$ symmetric criticality we obtain a larger 
correlation length critical exponent, $\nu^{O(4)}=2/3$, which at this order is independent of the CS level. Interestingly,  
the same value is obtained for the limit case of a neutral system. 


Finally, we would like to calculate the anomalous dimension $\eta_N$ of the critical 
magnetization correlation function. 
Using the relation, 
\begin{equation}
	\sigma^a_{\alpha\beta}\sigma^a_{\gamma\delta}=2\delta_{\alpha\delta}\delta_{\beta\gamma}
	-\delta_{\alpha\beta}\delta_{\gamma\delta}, 
\end{equation}
we obtain that, 
\begin{eqnarray}
	\label{Eq:nn-correlation}
	\mathcal{G}(x)=\langle\vec{n}(x)\cdot\vec{n}(0)\rangle&=&2\langle z_\alpha^*(x)z_\beta(x)z_\alpha(0)
	z_\beta^*(0)\rangle
	\nonumber\\
	&-&\langle|z_\alpha(x)|^2|z_\beta(0)|^2\rangle. 
\end{eqnarray}
Note that in the hard constraint case the second term in the equation above is unity. 

The calculation of $\eta_N$ amounts to finding the anomalous dimension of the 
operator $z_\alpha^*(x)z_\beta(x)$ \cite{zinn2002quantum}. The anomalous 
dimension of this operator is one of the eigenvalues occurring in the matrix, 
\begin{equation}
	\llbracket\eta_2\rrbracket=-\mathcal{N}\frac{\hat{U}_*}{16\pi}\llbracket P\rrbracket
	-\left[\frac{\hat{U}_*}{8\pi}-\frac{2\hat{e}_*^2}{3\pi(1+\hat{e}_*^2|\kappa|)^2}
	\right]\llbracket I\rrbracket, 
\end{equation}  
where $\mathcal{N}=2$ corresponds to the $O(2)\times O(2)$ symmetric case and $\mathcal{N}=4$ to the 
$O(4)$ symmetric one. The matrix elements of $\llbracket P\rrbracket$ and 
$\llbracket I\rrbracket$ are given by, 
\begin{equation}
	\llbracket I\rrbracket_{\alpha\beta,\gamma\delta}=\frac{1}{2}(\delta_{\alpha\gamma}\delta_{\beta\delta}+\delta_{\alpha\delta}\delta_{\beta\gamma}),
\end{equation}
\begin{equation}
	\llbracket P\rrbracket_{\alpha\beta,\gamma\delta}=\frac{1}{\mathcal{N}}\delta_{\alpha\beta}\delta_{\gamma\delta},
\end{equation}
where the indices run from 1 to 4. The trace of $\llbracket\eta_2\rrbracket$ yields 
$\eta_2\delta_{\alpha\beta}$, which corresponds to insertions of $|z_\alpha|^2$. In 
other words, we recover the formula for $1/\nu$ via the well known relation 
$\eta_2=1/\nu-2$. The insertion of the operator $z_\alpha^*(x)z_\beta(x)$ with 
$\alpha\neq \beta$ corresponds to the zero eigenvalue of $\llbracket P\rrbracket$, and so 
we obtain, 
\begin{equation}
	\widetilde{\eta}_2=-\frac{\hat{U}_*}{8\pi}+\frac{2\hat{e}_*^2}{3\pi(1+\hat{e}_*^2|\kappa|)^2},
\end{equation}
which is related to $\eta_N$ by the formula $\eta_N=1-2\widetilde{\eta}_2$. In the 
$O(2)\times O(2)$ symmetric case we obtain, 
\begin{equation}
	\eta_N^{O(2)\times O(2)}=\frac{1}{5}\left[7-\frac{48}{(1+6n)^2}\right],
\end{equation}
and similarly for $O(4)$ symmetry,
\begin{equation}
	\eta_N^{O(4)}=\frac{4}{3}\left[1-\frac{56}{5(1+6n)^2}\right]. 
\end{equation}
For a level 1 CS term we obtain, $	\eta_N^{O(2)\times O(2)}=59/49\approx 1.2$ and 
$\eta_N^{O(4)}=164/147\approx 1.12$.

\section{Controlling the RG analysis}
\label{sec:control}

The advantage of the $\epsilon$-expansion is that it allows for a reliable expansion parameter in perturbation theory. Of course, one then is ultimately interested in the $\epsilon=1$ case, and several mathematical techniques have been used in the past to show that the perturbation series actually converge \cite{zinn2002quantum}. There is also the fixed dimension approach by Parisi \cite{Parisi}, but this typically applies to scalar field theories without the coupling to a gauge field. The difficulty can be seen in our case, where for $N=2$, which is the case we are interested in, the RG fixed point $\hat{e}^2_*=12\pi$ is too large. We cannot use the $\epsilon$-expansion to obtain $\hat{e}^2_*\sim\mathcal{O}(\epsilon)$ in this case because the CS term imposes a fixed dimension $d=3$ from the outset. Furthermore, thanks to the CS term the Feynman diagram of Fig. \ref{Fig:photon4p-loop} vanishes. This diagram is a known obstruction towards reaching a fixed point for the scalar couplings, since it leads to a large contribution in the $\beta$ functions, even within the $\epsilon$-expansion. Only for $N$ sufficiently large the theory can become critical for a nonzero gauge coupling \cite{Hikami}. Hence, fixed dimensionality $d=3$ is a desirable feature in our case. 

Introducing a larger global symmetry group at fixed dimension $d=3$ provides a way to control the perturbation expansion, since all fixed points behave as $\sim\mathcal{O}(1/N)$ for $N$ large.  However, let us give an additional argument that even though we take $N=2$ at the end, the results for universal quantities can be relied upon.

The main source of difficulty for $N=2$ is the fixed point value for the gauge coupling, which is $\hat{e}^2_*=12\pi$ in this case. However, quite generally, the renormalization of the gauge coupling follows from the self-energy of the gauge field propagator, which is obtained from the vacuum polarization diagram of Fig. \ref{Fig:poldiag}. This leads to the effective Maxwell Lagrangian, 
\begin{equation}
	\label{Eq:Maxwell-eff}
	\mathcal{L}_{M}=\frac{1}{2e^2}\left[1+\Pi(0)\right](\epsilon_{\mu\nu\lambda}\partial_\nu a_\lambda)^2,
\end{equation} 
where $\Pi(0)=Ne^2/(24\pi m)$ is the vacuum polarization at $p=0$. 
From this we read off the dimensionless renormalized gauge coupling, 
\begin{eqnarray}
	\label{Eq:e2-eff}
	\hat{e}^2&=&\frac{e^2m^{-1}}{1+\Pi(0)}
	=\frac{e^2m^{-1}}{1+\frac{Ne^2}{24\pi m}}.
\end{eqnarray}
Thus, the $\beta$ function of Eq. (\ref{Eq:betae2}) is easily obtained by simple differentiation, $\beta_{\hat{e}^2}=md\hat{e}^2/dm$ without any need of a series expansion. Furthermore, we note the following two important facts. First, the result obtained in Eq. (\ref{Eq:e2-eff}) is the same as the one obtained within an $1/N$ expansion. Second, the fixed point follows from Eq. (\ref{Eq:e2-eff}) in two different ways, namely, either by directly letting $m\to 0$, or by taking the limit where the bare (dimensionful) gauge coupling $e^2\to\infty$. The latter limit highlights the strong coupling character of the theory at $d=3$. Furthermore, this is the regime of interest to us in the duality analysis. 

As far as the couplings $\hat{U}$ and $\hat{V}$ are concerned, $\hat{e}^2$ enters only via the wavefunction renormalization, since the diagram of Fig. \ref{Fig:photon4p-loop} vanishes. Since the CS mass also depends on $e^2$, the perturbative results in Eqs. (\ref{Eq:U-exp}) and (\ref{Eq:V-exp}) are not jeopardized by the strong-coupling character of the gauge coupling.     

Finally, we could, somewhat artificially, make an $\epsilon$-expansion analysis in which we compute Feynman diagrams in $d$ dimensions for the cases where  $\epsilon_{\mu\nu\lambda}$ does not play any role, while still keeping $d=3$ in the diagram of Fig. \ref{Fig:photon4p-loop}, since in this case $\epsilon_{\mu\nu\lambda}$ plays a crucial role. It is worth to carry out this calculation as well, in order to clearly show the need of the fixed dimension approach in this case. In fact, we will show below that while fixed points exist as before, they lead to unphysical values of the critical exponent $\nu$ in the $O(2)\times O(2)$ invariant case. 

Most of what we need for this calculation is already available, since we have discussed the $d$-dimensional example in absence of the gauge coupling earlier in the previous section. It remains to discuss the changes in the diagram of Fig. \ref{Fig:self-energy}. We have, 
\begin{eqnarray}
	\Sigma(p)&=&\frac{4e^2m^{d-4}}{(4\pi)^{d/2}}\left(1-\frac{1}{d}\right)\Gamma\left(1-\frac{d}{2}\right)
	\nonumber\\
	&\times&\frac{1-(m/M)^{d-2}}{1-M^2/m^2}p^2+\mathcal{O}(p^4),
\end{eqnarray} 
which upon expanding around $d=4$ yields the wavefunction renormalization at one-loop order, 
\begin{equation}
	Z=1+\frac{3\hat{e}^2}{8\pi^2\epsilon}. 
\end{equation}
Hence, the $\beta$ functions become for $N=2$, 
\begin{equation}
	\label{Eq:beta-U-2}
	\beta_{\hat{U}}=-(\epsilon+6\hat{e}^2)\hat{U}+5\hat{U}^2+\hat{V}^2,
\end{equation}
\begin{equation}
	\label{Eq:beta-V-2}
	\beta_{\hat{V}}=-(\epsilon+6\hat{e}^2)\hat{V}+2(\hat{V}^2+2\hat{U}\hat{V}),
\end{equation}
\begin{equation}
	\label{Eq:beta-e2}
	\beta_{\hat{e}^2}=-\epsilon\hat{e}^2+\frac{\hat{e}^4}{3}, 
\end{equation}
\begin{equation}
	\beta_\kappa=\left(\frac{\hat{e}^4}{3}-\epsilon\right)\kappa,
\end{equation}
where we have performed a rescaling similar to the one described above Eqs. (\ref{Eq:beta-U}) and (\ref{Eq:beta-V}). 

Now, if we consider the correlation length exponent for the $O(2)\times O(2)$ case, we obtain to order $\epsilon$, 
\begin{equation}
	\label{Eq:nu-unphys}
	\nu_{O(2)\times O(2)}=\frac{1}{2+8\epsilon/3}\approx\frac{1}{2}-\frac{2\epsilon}{3},
\end{equation}
and we see after setting $\epsilon=1$ at the end that $\nu_{O(2)\times O(2)}<0$ and therefore unphysical. Even if one does not completely adhere to the $\epsilon$-expansion and use Eq. (\ref{Eq:nu-unphys}) without making the expansion, a result smaller than 1/2 is obtained after setting $\epsilon=1$. This is also unphysical, since the critical exponent $\nu$ should be larger than or equal to its mean-field value for a local field theory of this type.

\section{Self-duality}

Let us introduce a change of the variables for the gauge fields, $b_{+\mu} = (b_{1\mu}+b_{2\mu})/2$, $b_{-\mu} = (b_{1\mu}-b_{2\mu})/2$. Then, the Eq. (13) of the paper takes the form,
\begin{eqnarray}
	\label{Eq:Seff-dual}
	\mathcal{L}_{\text{dual}}&=&g[(\epsilon_{\mu\nu\lambda}\partial_\nu b_{+ \lambda})^2 +(\epsilon_{\mu\nu\lambda}\partial_\nu b_{- \lambda})^2 ]\nonumber\\	
	&+&i2\pi (w_{1\mu} +w_{2\mu}) b_{+\mu}+  i2\pi (w_{1\mu} -w_{2\mu}) b_{-\mu}\nonumber\\&-&\frac{i}{2\kappa}\epsilon_{\mu\nu\lambda}
	b_{+\mu} \partial_\nu b_{+\lambda}.
\end{eqnarray}
To integrate out the gauge fields, we need to find the propagator $\widetilde{D}_{\mu \nu} $ which is the inverse of the tensor,
\begin{equation}
	M_{\mu \nu} =2 g\left(p^{2} \delta_{\mu \nu}-p_{\mu} p_{\nu}\right)+\frac{2 }{\alpha} p_{\mu} p_{\nu}
	+\frac{4}{\kappa} \varepsilon_{\mu \nu \lambda} p_{\lambda},
\end{equation}
with a gauge fixing $\alpha$. 
After a straightforward calculation one obtains the propagator $\widetilde{D}_{\mu \nu}$ in the momentum space,
\begin{equation}
	\widetilde{D}_{\mu \nu}(p)=\frac{1}{2\left(g^{2} \kappa^{2} p^{2}+4\right)}\left(\delta_{\mu \nu}-2\frac{\varepsilon_{\mu \nu \lambda}p_{\lambda}}{\kappa g p^{2}}\right),
	\label{eq:propagator}
\end{equation}
where the Landau gauge ($\alpha =0$) was used and we have dropped a longitudinal part $\sim p_\mu p_\nu$, since the zero divergence constraint of the vortex loop variables causes such a term to give a vanishing contribution. Therefore, the effective action for vortex fields in the momentum space is
\begin{eqnarray}
	\label{eq:sdual-momentum-dmunu}
	S_{\text {dual }}&=&2 \pi^{2} g \kappa^{2} \int_{p} D_{\mu \nu}(p)(w_{1 \mu}+w_{2 \mu} )(p)(w_{1 \mu}+w_{2 \mu})(-p) \nonumber\\
	&+&\frac{\pi^{2}}{g} \int_{p} \frac{(w_{1 \mu}-w_{2 \mu} )(p)(w_{1 \mu}-w_{2 \mu})(-p)}{p^{2}}.
\end{eqnarray}
And in the coordinate space one obtains,
\begin{eqnarray}
	\label{eq:sdual-coor-dmunu}
	S_{\text {dual }}&=&2 \pi^{2} g \kappa^{2} \int_{x} \int_{x'} D_{\mu \nu}(x-x')(w_{1 \mu}+w_{2 \mu} )(w_{1 \mu}'+w_{2 \mu}')\nonumber\\
	&+&\frac{\pi^{2}}{g} \int_{x} \int_{x'}\frac{(w_{1 \mu}-w_{2 \mu} )(w_{1 \mu}'-w_{2 \mu}')}{|x-x'|}.
\end{eqnarray}
We perform explicit calculations with the propagator Eq. \eqref{eq:propagator}, the first term in Eq. \eqref{eq:sdual-momentum-dmunu} becomes
\begin{eqnarray}
	&&2 \pi^{2} g \kappa^{2} \int_{p} D_{\mu \nu}(p)(w_{1 \mu}+w_{2 \mu} )(p)(w_{1 \mu}+w_{2 \mu})(-p)\nonumber\\&=& 2 \pi^{2} g \kappa^{2} \left( \int_{p} \frac{(w_{1 \mu}+w_{2 \mu} )(p)(w_{1 \mu}+w_{2 \mu})(-p)}{2\left(g^{2} \kappa^{2} p^{2}+4\right)}\right.\nonumber\\&-&\left.  \int_{p} \frac{\varepsilon_{\mu \nu \lambda}p_{\lambda}(w_{1 \mu}+w_{2 \mu} )(p)(w_{1 \nu}+w_{2 \nu})(-p)}{\kappa g p^2(g^{2} \kappa^{2} p^{2}+4)}\right),
\end{eqnarray}
In the case of $g^2p^2\ll 1$, the last line of the expression simplifies. Let us take a closer look at the integral not taking coefficients into account,
\begin{eqnarray}
	&& \int_{p} \frac{\varepsilon_{\mu \nu \lambda}p_{\lambda}(w_{1 \mu}+w_{2 \mu} )(p)(w_{1 \nu}+w_{2 \nu})(-p)}{ p^2}\nonumber\\&=&  \int_{p}\ \frac{\varepsilon_{\mu \nu \lambda} p_{\lambda}}{p^{2}} \int_{x} \int_{y} e^{i p\cdot(y-x)} (w_{1\mu}+w_{2\mu})(x) (w_{1\nu}+w_{2\nu})(y)\nonumber\\&=& 
	\frac{i}{4\pi} \int_{p} \int_{z} \frac{\varepsilon_{\mu \nu \lambda} z_{\lambda} e^{-i p \cdot y}}{|z|^{3}} \nonumber \\ &\times&\int_{x} \int_{y} e^{i p \cdot(y - x)}(w_{1\mu}+w_{2\mu})(x) (w_{1\nu}+w_{2\nu})(y), 
\end{eqnarray}
where we used a Fourier transform and an exponential representation of the $\delta$-function. Further calculations lead to
\begin{eqnarray}
	&&\frac{i}{4\pi}  \varepsilon_{\mu \nu \lambda} \int_{x} \int_{y} \frac{\left(x_{\lambda}-y_{\lambda}\right)}{|x-y|^{3}}(w_{1\mu}+w_{2\mu})(x) (w_{1\nu}+w_{2\nu})(y) \nonumber \\&=&-\frac{i}{4\pi} \int_{x} \int_{y}\frac{\left(x_{\alpha}-y_{\alpha}\right)}{|x-y|^{3}} \partial_{\alpha} (v_{1\beta}+v_{2\beta})(x) (w_{1\beta}+w_{2\beta})(y)\nonumber \\&=& i \int d^3x (v_{1\beta}+v_{2\beta}) (w_{1\beta}+w_{2\beta}).
\end{eqnarray}
Finally, the Eq. \eqref{eq:sdual-coor-dmunu} takes the form,
\begin{eqnarray}
	\label{eq:sdual}
	S_{\text {dual }}&=& \frac{\pi}{4 g} \int_{x} \int_{x'} e^{-\frac{2|x-x'|}{g\kappa}}\frac{(w_{1 \mu}+w_{2 \mu} )(x)(w_{1 \mu}+w_{2 \mu})(x')}{ |x-x'|} \nonumber\\
	&+&\frac{\pi^{2}}{g} \int_{x} \int_{x'}\frac{(w_{1 \mu}-w_{2 \mu} )(w_{1 \mu}'-w_{2 \mu}')}{|x-x'|}\nonumber\\
	&+&\frac{i\pi^{2}\kappa }{2} \int  d^3x (v_{1\beta}+v_{2\beta}) (w_{1\beta}+w_{2\beta}).
\end{eqnarray}

\section{Flux attachment bosonization duality}
In this section we conciser the bosonization duality that we obtain in the scope of a duality web approach \cite{Karch_PhysRevX.6.031043,SEIBERG2016395}. To do so, we first  need to write down the field theory for the dual bosonic system obtained in Eq. (13) of the main body of the paper. To this end we introduce complex scalar fields $\phi_I$, $I=1,2$ yielding a second-quantized representation for the ensemble of vortex loops \cite{kleinert1989gauge}. This yields the Lagrangian,  
\begin{eqnarray}
	\label{Eq:Dual-L}
	\mathcal{L}_{dual}&=&\sum_{I=1,2}\left[|(\partial_\mu-ib_{I\mu})\phi_I|^2+m^2|\phi_I|^2+\frac{\lambda}{2}|\phi_I|^4
	\right]\nonumber\\
	&-&\frac{i}{8\pi^2\kappa}\epsilon_{\mu\nu\lambda}(b_{1\mu}+b_{2\mu})\partial_\nu (b_{1\lambda}+ b_{2\lambda}),
\end{eqnarray}
where we have rescaled gauge fields $b_{I\mu}\to b_{I\mu}/(2\pi)$, thus assigning a unit charge to both $\phi_1$ and $\phi_2$. Following the technique employed in Ref.  \cite{Karch_PhysRevX.6.031043}, we awoke the bosonization conjectures,
\begin{equation}
	\label{Eq:Boson+flux}
	Z_{f Q E D}[A] e^{\frac{1}{2}S_{CS}[A]}=Z_{b Q E D+flux}[A],
\end{equation}
\begin{equation}
	\label{Eq:Fermion+flux}
	Z_{f Q E D+flux}[A]=Z_{b Q E D}[A] e^{-S_{CS}[A]},
\end{equation}
where $S_{CS}[A]$ is the action for a level 1 CS term, $A_\mu$ is the background field. The fermionic and bosonic partition functions are, respectively,
\begin{eqnarray}
	\label{Eq:ZfQED}
	Z_{fQED}[A]&=&\int\mathcal{D}\bar{\psi}\mathcal{D}\psi e^{-S_{fQED}[A]},
	\nonumber\\
	S_{fQED}[A]&=&\int d^3x\bar{\psi}(\slashchar{\partial}-i\slashchar{A})\psi ,
\end{eqnarray}
\begin{eqnarray}
	\label{Eq:ZbQED}
	Z_{bQED}[A]&=&\int\mathcal{D}\phi^*\mathcal{D}\phi e^{-S_{bQED}[A]},
	\nonumber\\
	S_{bQED}[A]&=&\int d^3x\left[|(\partial_\mu-iA_\mu)\phi|^2+m^2|\phi|^2+\frac{\lambda}{2}|\phi|^4
	\right],\nonumber\\
\end{eqnarray}
while the flux attachment operates as follows, 
\begin{eqnarray}
	\label{Eq:ZfQED+flux}
	Z_{fQED+flux}[A]&=&\int\mathcal{D}a_\mu\mathcal{D}\bar{\psi}\mathcal{D}\psi e^{-S_{fQED+flux}[A]},
	\nonumber\\
	S_{fQED+flux}[A]&=&S_{fQED}[a] -\frac{1}{2}S_{CS}[a] - S_{BF}[a;A]
	\nonumber\\
\end{eqnarray}
\begin{eqnarray}
	\label{Eq:ZbQED+flux}
	Z_{bQED+flux}[A]&=&\int\mathcal{D}a_\mu\mathcal{D}\phi^*\mathcal{D}\phi e^{-S_{bQED+flux}[A]},
	\nonumber\\
	S_{bQED+flux}[A]&=&S_{bQED}[a] + S_{CS}[a] + S_{BF}[a;A]\nonumber\\
\end{eqnarray}
where the BF term is given by, 
\begin{equation}
	S_{BF}[a;A]=\frac{i}{2\pi}\int d^3 xa_\mu\epsilon_{\mu\nu\lambda}\partial_\nu A_\lambda. 
\end{equation}

Now that we have recalled the basic flux attachment dualities (\ref{Eq:Boson+flux}) and (\ref{Eq:Fermion+flux}), we can derive the duality described in the main text by multiplying both these relations together, 
\begin{eqnarray}
	\label{eq:zmult}
	&Z&_{f Q E D}[A] e^{\frac{1}{2}S_{CS}[A]} 	Z_{f Q E D+flux}[A]\nonumber\\=&Z&_{b Q E D+flux}[A]Z_{b Q E D}[A] e^{-S_{CS[A]}},
\end{eqnarray}
After we promote the background field $A_\mu$ to a dynamical field $b_\mu$, the left-hand side of the Eq. \eqref{eq:zmult} takes the form,
\begin{eqnarray}
	&&\int\prod_{I=1,2}\mathcal{D}b_{I\mu}\mathcal{D}\bar{\psi}_I\mathcal{D}\psi_I e^{-S},
	\nonumber\\
	S&=&\int d^3x\left\{\sum_{I=1,2}\left[ \bar{\psi}_I(\slashchar{\partial}-i\slashchar{b}_I)\psi_I-\frac{i}{8\pi}b_{I\mu}\epsilon_{\mu\nu\lambda}\partial_\nu b_{I\lambda}\right] \right.\nonumber\\
	&-& \left.\frac{i}{2\pi}b_{1\mu}\epsilon_{\mu\nu\lambda}\partial_\nu b_{2\lambda}\right\}.\nonumber\\
\end{eqnarray}
Integrating out $\psi_2$ generates a level 1/2 CS term with a minus sign. We can integrate out the dynamical field $b_{2\mu}$, which enforces $b_{1\mu}=b_\mu$. Eventually, we can write down the left-hand side of the Eq. \eqref{eq:zmult},
\begin{eqnarray}
	&&\int\mathcal{D}b_{\mu}\mathcal{D}\bar{\psi}\mathcal{D}\psi e^{-\int d^3x\left[\bar{\psi}(\slashchar{\partial}-i\slashchar{b})\psi + \frac{i}{8\pi}b_{\mu}\epsilon_{\mu\nu\lambda}\partial_\nu b_{\lambda}\right]},
\end{eqnarray}
where we have set $\psi_{1}\equiv \psi$.

Now, let us write explicitly the right-hand side of the  Eq. \eqref{eq:zmult},
\begin{eqnarray}
	&&\int\prod_{I=1,2}\mathcal{D}b_{I\mu}\mathcal{D}\phi^*_I\mathcal{D}\phi_I e^{-S},
	\nonumber\\
	S&=&\int d^3x\left\{\sum_{I=1,2}\left[ |(\partial_\mu-ib_{I\mu})\phi_I|^2+… 
	\right] \right.\nonumber\\
	&+& \left.
	\frac{i}{4\pi}\epsilon_{\mu\nu\lambda}(b_{1\mu}+b_{2\mu})\partial_\nu (b_{1\lambda}+ b_{2\lambda})
	\right\},
	\nonumber\\
\end{eqnarray}
where the ellipsis represent scalar field self-interactions. This is precisely the time-reversal transformed version of the dual Lagrangian (\ref{Eq:Dual-L}) for $\kappa=1/(2\pi)$. Therefore, we have obtained that our derivation is consistent with the bosonization duality performed via flux attachments to fermions and bosons. 

\end{document}